\documentclass[12pt]{article}

\usepackage{graphics,amssymb,float}
\usepackage{graphicx}
\usepackage{caption}
\usepackage{rotating}
\usepackage{dcolumn}
\usepackage{bm}
\usepackage{cite}
\usepackage{amsmath}
\usepackage{indentfirst} 
\usepackage{epstopdf}
\usepackage[usenames,dvipsnames]{xcolor}
\usepackage{setspace}

\makeatletter

\@addtoreset{equation}{section}
\makeatother

\def\lsim{\;\raise0.3ex\hbox{$<$\kern-0.75em\raise-1.1ex\hbox{$\sim$}}\;}
\def\gsim{\;\raise0.3ex\hbox{$>$\kern-0.75em\raise-1.1ex\hbox{$\sim$}}\;}
\def\beq{\begin{equation}}   \def\eeq{\end{equation}}
\def\ba{\begin{array}}       \def\ea{\end{array}}
\def\bea{\begin{eqnarray}}   \def\eea{\end{eqnarray}}
\def\nn{\nonumber}

\begin{document}

\begin{titlepage}
\begin{flushright}
LUPM 22-004\\
\end{flushright}


\begin{center}
\vspace{1cm}

{\Large\bf Benchmark Planes for Higgs-to-Higgs Decays in the}

\vspace{2mm}
{\Large\bf NMSSM} 

\vspace{2cm}

{\bf{Ulrich Ellwanger$^a$ and Cyril Hugonie$^b$}}\\
\vspace{1cm}
\it $^a$ IJCLab,  CNRS/IN2P3, University  Paris-Saclay, 91405  Orsay,  France\\
ulrich.ellwanger@ijclab.in2p3.fr\\
\it $^b$ LUPM, UMR 5299, CNRS/IN2P3, Universit\'e de Montpellier, 34095 Montpellier, France\\
cyril.hugonie@umontpellier.fr

\end{center}
\vspace{2cm}

\begin{abstract}
We present benchmark planes (or lines) with cross sections via gluon fusion for the processes $H\to h+H_{S}$, resonant Higgs pair and triple Higgs production, $A \to h+(A_S \to \gamma\gamma)$,
$A \to Z+ h$ and $A \to Z+ H_{S}$  within the Next-to-Minimal Supersymmetric Standard Model. 
Moreover we propose new searches for $H\to  h+(H_{S} \to t\bar{t})$, $A \to Z+ (H_{S}\to h+h)$ and $A \to Z+ (H_{S}\to t\bar{t})$ for which possible cross sections are given. These allow the experimental collaborations to verify in the future which search channels cover yet unexplored regions of the parameter space. Expressions for the dominant contributions to trilinear Higgs couplings and Higgs--Z couplings are discussed which allow to identify the dominant processes contributing to a given final state.
\end{abstract}

\end{titlepage}

\section{Introduction}

Searches for new particles beyond the Standard Model (SM) at the LHC are difficult if light new particles have small production cross sections, and heavy new particles with larger production cross sections undergo dominantly cascade decays. This can be the case for supersymmetric extensions of the SM such as the Next-to-Minimal Supersymmetric Standard Model (NMSSM) \cite{Maniatis:2009re,Ellwanger:2009dp}. In the present article we focus on the Higgs sector of the NMSSM, where this phenomenon can take place. Accordingly wide regions of the parameter space of the Higgs sector of the NMSSM have not yet been explored, and it remains a challenging task for the future to cover them.

The ATLAS and CMS experiments at the LHC have started to search for final states resulting from one-step cascade decays in Higgs sectors beyond the SM. Given the existing phenomenological constraints from previous searches \cite{CMS:2018rmh,CMS:2018amk,CMS:2019mij,CMS:2019pzc,ATLAS:2020zms,ATLAS:2020tlo,ATLAS:2021uiz,CMS:2021klu,CMS:2021yci,CMS-PAS-B2G-21-003,CMS:2019ogx,ATLAS:2020gxx,CMS:2019qcx,CMS:2019kca,ATLAS:2020pgp,CMS:2022qww,ATLAS:2018ili,CMS:2018tla,CMS:2018qmt,CMS:2018vjd,ATLAS:2018fpd,CMS:2018ipl,CMS:2020jeo,ATLAS:2021ifb,ATLAS:2021tyg,CMS:2021roc,Veatch:2022uzz}, from properties of the mostly SM-like Higgs boson at 125~GeV \cite{ATLAS:2016neq,CMS:2018uag,ATLAS:2020qdt}, from direct detection experiments of dark matter
\cite{CRESST:2015txj,DarkSide:2018bpj,XENON:2019rxp,PICO:2019vsc}
 and more, it is a priori not clear in how far future such searches will explore new regions in the parameter space of the NMSSM. To this end it is useful to map out benchmark planes or lines satisfying existing phenomenological constraints, which is the task of the present article.

The Higgs sector of the NMSSM consists in two SU(2) doublets and a complex SU(2) singlet. In the CP-conserving NMSSM, the physical scalars can be decomposed into three neutral CP-even states, two neutral CP-odd states and one complex charged state. One of the three neutral CP-even states has to correspond to the SM-like Higgs boson. A priori the masses of the remaining states can assume a large range of values, depending on the five NMSSM-specific parameters $\lambda$, $\kappa$, $A_\lambda$, $A_\kappa$, $\mu_{\rm eff}$ as well as on $\tan\beta$ \cite{Maniatis:2009re,Ellwanger:2009dp}. 

In general, the three neutral CP-even states as well as the two neutral CP-odd states are mixtures of SU(2) doublets and a SU(2) singlet. Thereby all scalars obtain couplings to SM fermions and gauge bosons (originally reserved to the SU(2) doublets), and all possible CP-conserving trilinear couplings among CP-even and CP-odd scalars are non-zero. Still, in most of the parameter space one can denote each of the three CP-even scalars $H_1$, $H_2$ and $H_3$ (ordered in mass) as either SM-like ($h$), or mostly ``singlet-like'' ($H_S$, with small direct production cross sections) or mostly ``MSSM-like'' ($H$). Here MSSM-like refers to a nearly degenerate SU(2) doublet (if much heavier than the SM-like Higgs boson) consisting in a neutral CP-even, a neutral CP-odd and a charged complex state. Likewise one can denote each of the two CP-odd scalars $A_1$, $A_2$ as either mostly singlet-like ($A_S$) or mostly MSSM-like ($A$).

Given lower bounds on the mass of the MSSM-like charged Higgs boson both from direct searches and from $b \to s+\gamma$ (although the charged Higgs contribution can partially be cancelled by supersymmetric contributions), the neutral CP-even and CP-odd members of the MSSM-like SU(2) doublet cannot be light; 350~GeV is a conservative lower bound on their masses. In contrast, any range is still allowed for the (independent) masses of the mostly singlet-like CP-even and CP-odd states. However, if lighter than $~60$~GeV, their couplings to the SM-like Higgs boson must be small enough in order to escape bounds from searches for decays of the SM-like Higgs boson into pairs of light scalars, and bounds from beyond-the-SM contributions to the total width of the SM-like Higgs boson.

Searches for heavier neutral scalars such as the neutral CP-even and CP-odd members of the MSSM-like SU(2) doublet can focus on their production via gluon fusion (ggF) and their decays into pairs of fermions or gauge bosons leading to events with little SM background once resonance-like excesses are looked for \cite{CMS:2018rmh,CMS:2018amk,CMS:2019mij,CMS:2019pzc,ATLAS:2020zms,ATLAS:2020tlo,ATLAS:2021uiz,CMS:2021klu}.

However, in models with extended Higgs sectors such as the NMSSM, heavy scalars can have sizeable branching fractions into two lighter scalars, or a $Z$~boson and one scalar. For scalars with small direct production cross sections, such processes can be the only way to discover them. Corresponding searches have been performed in \cite{CMS:2021yci,CMS-PAS-B2G-21-003,CMS:2019ogx,ATLAS:2020gxx,CMS:2019qcx,CMS:2019kca,ATLAS:2020pgp,CMS:2022qww}.
Finally heavy CP-even scalars can also be searched for in final states corresponding to resonant SM Higgs pair production, see 
\cite{ATLAS:2018ili,CMS:2018tla,CMS:2018qmt,CMS:2018vjd,ATLAS:2018fpd,CMS:2018ipl,CMS:2020jeo,ATLAS:2021ifb,ATLAS:2021tyg,CMS:2021roc}.
For a recent review of boson pair production at the LHC see \cite{Veatch:2022uzz}.

Benchmark points for Higgs-to-Higgs cascade processes in the NMSSM have been proposed in 
\cite{Kang:2013rj,King:2014xwa,Carena:2015moc,Ellwanger:2015uaz,Costa:2015llh,Baum:2017gbj,Ellwanger:2017skc,Baum:2017enm,Basler:2018dac,Baum:2019uzg,Baum:2019pqc,Barducci:2019xkq,Biekotter:2021qbc,Abouabid:2021yvw}, see also the twiki web site of the LHC-HXSWG3-NMSSM working group \cite{LHCHXSWG3NMSSM}.
However, earlier benchmark points are often outdated due to more recent limits from searches in the proposed (or other) channels.

The benchmark points presented here are chosen such that constraints from the existing searches above are satisfied. In addition we impose constraints from B-physics, constraints from properties of the SM-like Higgs boson (a mass within $125\pm 2$~GeV allowing for theoretical uncertainties, and couplings in the $\kappa$-framework satisfying combined limits of ATLAS and CMS \cite{ATLAS:2016neq,CMS:2018uag,ATLAS:2020qdt}), and constraints from stability of the electroweak vacuum. Constraints from the anomalous magnetic moment of the muon are left aside as these concern the smuon/gaugino sector which is irrelevant here. 

We also require that the lightest supersymmetric particle is neutral (the lightest neutralino), since it is stable and contributes necessarily to the relic density of the universe. We do not require that it accounts for {\it all} of the observed dark matter relic density as there may exist additional contributions from physics far above the weak scale. (For this reason we do not require the absence of a Landau singularity below the GUT scale but confine ourselves to $\lambda < 0.7$ in order to avoid a strong coupling regime close to the weak scale.)
However, the stable lightest neutralino unavoidably contributes to dark matter direct detection experiments, and must satisfy corresponding constraints which are imposed on the benchmark points since the properties of the lightest neutralino (mass and annihilation rate typically via a CP-even or CP-odd scalar in the s-channel) depend on the same parameters as the NMSSM Higgs sector.

The above constraints are implemented in the code NMSSMTools\_5.6.2 \cite{Ellwanger:2004xm,Ellwanger:2005dv} (for more details see the website \cite{NMSSMTools}) coupled to MicrOmegas \cite{Belanger:2013oya} for the calculation of the dark matter relic density and direct detection cross sections.

Usually the production cross section of heavy Higgs states via gluon fusion dominates and is considered here, although vector boson fusion can be relevant in some particular regions of the parameter space \cite{Das:2018fog}.
For the calculation of the cross sections $ggF\to H/A$ (with $M_{H/A} \geq 400$~GeV) we start with the BSM Higgs production cross sections at $\sqrt{s} = 13$~TeV (update in CERN Report4 2016) from the twiki web page \cite{LHCyellowreport}. These are multiplied by the reduced couplings squared of $H/A$. Thereby we capture most of the radiative QCD corrections in the form of K-factors; the remaining theoretical uncertainties are at most of ${\cal O}(10\%) $.

In the next Section~2 we discuss masses and trilinear couplings (including $H_i$--$A_j$--Z) in the Higgs basis in the NMSSM confining ourselves to numerically dominant contributions. This allows to estimate which cascade decays are usually dominant. In Section~3 we present benchmark planes for various final states corresponding to $H\to h+H_{S}$ in the space $M_{H} - M_{H_{S}}$. For $M_{H_{S}}$ we confine ourselves to the range $M_{H_{S}}>60$~GeV: Otherwise the couplings of $H_{S}$ must be small enough in order to satisfy constraints from $h\to H_{S} + H_{S}$ leading to small allowed cross sections for its production via cascade decays, and its discovery seems more likely via decays of $h$. We also present a benchmark line with the largest possible cross sections for resonant SM-Higgs pair production $H_2\to h+h$ as function of $M_{H_2}$ ($H_2$ is a mixture of $H_S$ and $H$), cross sections for triple SM-Higgs production and for the yet unexplored process $H\to h + (H_{S}\to t\bar{t})$. 

In the NMSSM, singlet-like pseudoscalars $A_S$ can have dominant branching fractions into $\gamma\gamma$: tree level couplings to SM gauge bosons and fermions can vanish, but couplings to higgsino-like charginos $\sim \lambda$ remain. While decays into chargino pairs are kinematically forbidden, chargino loops induce a coupling of $A_S$ to photons making the diphoton channel the dominant decay mode. (The branching fraction into the loop induced $Z+\gamma$ channel is about half as large unless kinematically suppressed, in which case the branching fractions into $\gamma\gamma$ can become 99\%.) The production of $A_S$ can proceed via the production of the MSSM-like pseudoscalar $A$, and its decay $A\to A_S + h$. Benchmark points for this process will be given as well.

Furthermore we show benchmark planes for final states corresponding to $A\to H_{S}+Z$ in the space $M_{A}-M_{H_{S}}$. These benchmark points are the same as for $H\to h+H_{S}$ which allows to compare the cross sections, and hence to estimate the corresponding relative sensitivities. Cross sections for the yet unexplored processes $A\to Z + (H_{S}\to h+h)$ and $A\to Z + (H_{S}\to t\bar{t})$ are also given.
Finally we present a benchmark line with the largest possible cross sections for $A\to h+Z$ as function of $M_{A}$.
A summary is given in Section~4.

\section{The Higgs Sector and Trilinear Higgs Couplings in the NMSSM}

The neutral Higgs sector of the NMSSM consists in three complex scalars $H_u^0$, $H_d^0$ and $S$ where $H_u^0$ and $H_d^0$ are members of SU(2) doublets and $S$ is a gauge singlet \cite{Maniatis:2009re,Ellwanger:2009dp}. Their self couplings originate from terms 
\beq
W=\lambda {\cal H}_u {\cal H}_d {\cal S} + \frac{\kappa}{3} {\cal S}^3 + \dots
\eeq
in the superpotential $W$ in terms of superfields ${\cal H}$ and ${\cal S}$, and from trilinear soft supersymmetry breaking terms
\beq
(\lambda A_\lambda  H_u  H_d  S + \frac{\kappa}{3} A_\kappa S^3) + \text{h. c.}\; ;
\eeq
contributions from D-terms are relatively small. The self couplings have to be expressed in terms of physical states. To this end the weak eigenstates $H_u^0$, $H_d^0$ and $S$  have to be expanded around their vacuum expectation values $v_u$, $v_d$ and $s$ (where $v^2=v_u^2+v_d^2 \simeq (174 \text{GeV})^2$, $M_Z^2=g^2 v^2$ with $g^2=\frac{g_1^2+g_2^2}{2}$). The mass matrices have to be diagonalized, and in the CP-conserving case one obtains three neutral CP-even scalars and two neutral CP-odd scalars (after elimination of the Goldstone boson). General expressions for these mass matrices including the dominant radiative corrections are given in \cite{Ellwanger:2009dp}. A first approximation to the physical states is obtained in the so-called Higgs basis where singlet-doublet mixing is neglected and the CP-even doublets are rotated by the same angle as the CP-odd sector. Defining $\tan\beta=\frac{v_u}{v_d}$ and using hats for the Higgs basis ($\widehat{H}_{SM}$ is near to but not yet exactly equal to the physical SM Higgs boson $h$, and $\widehat{H}$ is near to but not yet exactly equal to the physical MSSM-like Higgs boson $H$) one has
\begin{align}
H_d^0 & =\cos\beta \widehat{H}_{SM} + \sin\beta \widehat{H}\; , \quad  H_u^0 = \sin\beta \widehat{H}_{SM} - \cos\beta \widehat{H}\; ,\nn \\
 A_d & = \sin\beta \widehat{A}\; , \qquad  A_u  = \cos\beta \widehat{A}\; .
\end{align}
To these the pure singlet states $\widehat{H}_S$ and $\widehat{A}_S$ have to be added. The tree level elements of the $2\times 2$ mass matrix in the CP-odd sector in the basis $\widehat{A}, \widehat{A}_S$ are for the typical case $s, A_\lambda \gg M_Z$
\begin{align}
M^2_{A,11} &= \frac{2\lambda s(A_\lambda+\kappa s)}{\sin 2\beta}\; ,\nn \\
M^2_{A,22} &= -3\kappa A_\kappa s + {\cal O}(M_Z^2)\; ,\nn \\
M^2_{A,12} &= \lambda v (A_\lambda -2\kappa s)\; .
\label{MA}
\end{align}
The tree level elements of the $3\times 3$ mass matrix in the CP-even sector in the basis $\widehat{H}_{SM}, \widehat{H}, \widehat{H}_S$ are 
\begin{align}
M^2_{H,11} &= M_Z^2(\cos^2 2\beta + \frac{\lambda^2}{g^2}\sin^2 2\beta)\; , \nn \\
M^2_{H,22} &= \frac{2\lambda s(A_\lambda+\kappa s)}{\sin 2\beta} + {\cal O}(M_Z^2)\; ,\nn \\
M^2_{H,33} &= 2\kappa s(A_\lambda+4\kappa s)\; , \nn \\
M^2_{H,12} &=  {\cal O}(M_Z^2)\; ,\nn \\
M^2_{H,13} &= \lambda v (2\lambda s-(A_\lambda+2\kappa s)\sin 2\beta)\; ,\nn \\
M^2_{H,23} &= \lambda v(A_\lambda+2\kappa s)\cos 2\beta\; .
\label{MH}
\end{align}
Hence singlet-doublet mixing is of ${\cal O}(\frac{v}{s},\frac{v}{A_\lambda})$ relative to the diagonal elements, but can still be large if the corresponding diagonal elements are close to each other.

Trilinear couplings are proportional to one of the vacuum expectation values $v_u$, $v_d$ and $s$, or to one of the trilinear soft supersymmetry breaking terms $A_\lambda$ or $A_\kappa$. The latter contributes only to the trilinear singlet Higgs couplings which play a negligible role for Higgs-to-Higgs decays since pure singlets have tiny production cross sections.\footnote{The cross sections in Section~3 are derived including all couplings and dominant radiative corrections as included in NMSSMTools \cite{Ellwanger:2004xm,Ellwanger:2005dv,NMSSMTools}.}
General expressions for the trilinear couplings can be found in 
\cite{Ellwanger:2009dp}, but it is instructive to compare the ones relevant for Higgs-to-Higgs decays for the typical case $s, A_\lambda \gg v_u, v_d \approx M_Z$. In the Higgs basis these are (neglecting contributions of ${\cal O}(M_Z)$)
\begin{align}
a)&\sim \widehat{H} \widehat{H}_{SM} \widehat{H}_S: &\ &-\frac{\lambda}{\sqrt{2}} (2\kappa s + A_\lambda)\; , \nn \\
b)&\sim \widehat{H} \widehat{H}_{SM} \widehat{H}_{SM}: &\ &0\; , \nn \\
c)&\sim \widehat{H}_{S} \widehat{H}_{SM} \widehat{H}_{SM}: &\ &\frac{\lambda^2}{\sqrt{2}} s - \frac{\lambda}{\sqrt{2}}\sin\beta \cos\beta (2\kappa s + A_\lambda)\; , \nn \\
d)&\sim \widehat{H}_{S} \widehat{H} \widehat{H}: &\ &\frac{\lambda^2}{\sqrt{2}} s + \frac{\lambda}{\sqrt{2}}\sin\beta \cos\beta (2\kappa s + A_\lambda)\; , \nn \\
e)&\sim \widehat{A} \widehat{H}_{SM} \widehat{A}_S: &\ &\lambda{\sqrt{2}}\sin\beta \cos\beta (-2\kappa s + A_\lambda)\; , \nn \\
f)&\sim \widehat{A} \widehat{H} \widehat{A}_S: &\ &\frac{\lambda}{\sqrt{2}}(\sin^2\beta- \cos^2\beta) (-2\kappa s + A_\lambda)\; .
\label{trilin}
\end{align}

If the fields in the Higgs basis are good approximations to the physical fields,
relevant processes for searches for $ggF\to X \to Y + h$  are $ggF\to \widehat{H} \to \widehat{H}_S + \widehat{H}_{SM}$ (using the trilinear coupling a)) and $ggF\to \widehat{A} \to \widehat{A}_S + \widehat{H}_{SM}$ (using the trilinear coupling e)); singlet-like scalars have small production cross sections. The production cross sections for $\widehat{H}$ and $\widehat{A}$ are similar, but the trilinear couplings a) are larger than the trilinear couplings e) for $2\sin\beta \cos\beta < 1$ and/or cancellations in $(-2\kappa s + A_\lambda)$ for $\kappa s, A_\lambda > 0$ as considered here. This explains why (for similar masses of $\widehat{H}$ and $\widehat{A}$) the process $ggF\to \widehat{H} \to \widehat{H}_S + \widehat{H}_{SM}$ dominates over $ggF\to \widehat{A} \to \widehat{A}_S + \widehat{H}_{SM}$.

An exception is the final state $h+\gamma \gamma$ with $\gamma\gamma$ from a BSM scalar or pseudo-scalar. As stated in the introduction, the mostly singlet-like pseudoscalar $A_S$ can have a dominant branching fraction up to $\sim 99$\% into diphotons if $Z+\gamma$ is kinematically suppressed (up to $\sim 66\%$ otherwise). For maximal cross sections, radiative corrections from supersymmetric particles to the pseudo\-scalar mass matrix  play a relevant role.
In the next section we consider a benchmark plane for this process as well.
(A corresponding decoupling of $H_S$ does not happen since in the scalar sector two mixing angles would have to vanish simultaneously, which would require $\lambda s \to 0$ leading to massless higgsinos.)

At first sight the prospects for resonant SM Higgs pair production look dim: For $\widehat{H}$ with the largest production cross section via gluon fusion the dominant trilinear coupling b) vanishes, whereas $\widehat{H}_S$ would not be produced via gluon fusion. However the scalar fields in the Higgs basis are not necessarily close to physical fields, and $\widehat{H}$ and $\widehat{H}_{S}$ can strongly mix. Indeed we found that the cross sections for resonant SM Higgs pair production can be quite large (see the next section) in this case.

Next we turn to decays $H\to A+Z$ and $A\to H+Z$. The relevant couplings are
\beq
H_i(p) A_j(p') Z_\mu:\ -i g\, C_i^H C_j^A (p-p')_\mu
\eeq
where $C_i^H$ denote the $\widehat{H}$ components of the physical states $H_i$, and $C_j^A$ the $\widehat{A}$ components of the physical states $A_j$. 

Decays $H_3 \to A_2 +Z$ and $A_2 \to H_3 +Z$ are usually impossible for kinematic reasons if the physical states $H_3$, $A_2$ are well approximated by the Higgs basis $\widehat{H}$, $\widehat{A}$. 
Decays $H_3 \to A_1 +Z$ with $A_1 \sim A_{S}$ are proportional to the $\widehat{A}$ component of $ A_{S}$ induced by the off-diagonal element $M^2_{A,12}$ in \eqref{MA} which is usually quite small.

Decays $A_{2} \to H_{1,2} +Z$ are proportional to the $\widehat{H}$ components of $H_{1,2}$. These are typically larger for $H_{S}$ compared to $h$ since $M^2_{H,23} > M^2_{H,12}$ in \eqref{MH}. As a consequence cross sections for searches for $ggF\to H/A \to A/H +Z$ are usually dominated by $ggF\to A \to H_{S}+Z$ in the NMSSM. On the other hand searches for $ggF\to A \to h+Z$ are also frequently performed. In the NMSSM, the possible cross sections (for a given mass $M_A$) are also discussed in the next Section.

\section{Benchmark Planes and Lines}

A significant excess in final states corresponding to $H_3\to h+H_{2}$ would imply the simultaneous discovery of two new bosons beyond the Higgs sector of the Standard model, which may correspond to the CP-even scalars $H$ and $H_{S}$ in the NMSSM. 
(We recall that the physical states $H$ and $H_{S}$ are generally mixtures of the weak eigenstates.)
Based on an integrated luminosity of up to 140~fb$^{-1}$, corresponding searches have been performed by CMS in the channel $H_3\to (h\to \tau\tau)+ (H_{S}\to bb)$ for mass ranges $240\ \text{GeV} < M_{H_3} < 3000$~GeV and $60\ \text{GeV} < M_{H_S} <  2800$~GeV in \cite{CMS:2021yci}, and in the channel $H_3\to (h\to bb)+ (H_{S}\to bb)$ for mass ranges $900 \ \text{GeV} < M_{H_3} <  4000$~GeV and $60\ \text{GeV} < M_{H_S} <  600$~GeV in \cite{CMS-PAS-B2G-21-003}.

We have prepared a plane of viable benchmark points for $ggF\to H \to (H_{S} \to bb) + h$ covering the mass ranges $400\ \text{GeV} < M_{H} <  2000$~GeV, and $60\ \text{GeV} < M_{H_S} <  800$~GeV (or $M_{H_S} <M_{H}-200$~GeV).
Generally, masses of $H$ and $H_{S}$ are given with a precision of $\pm 0.5$~GeV, except for $M_{H_{S}}=60$~GeV which means $60\leq M_{H_{S}} \leq 60.5$~GeV such that constraints from CMS 
in \cite{CMS:2021yci}, valid for $60\leq M_{H_{S}}$, are satisfied.\footnote{There do exist viable regions with $M_{H_{S}}<60$~GeV in the NMSSM, with small couplings to $h$ in order to satisfy constraints on $h\to H_{S} + H_{S}$. But then cross sections for cascade decays into $H_{S}$ are very small as well and are omitted here.}
Details of the NMSSM-specific parameters, masses, branching fractions and more for each point can be obtained in SLHA format from the authors.

For given values of $M_{H}$ and $M_{H_{S}}$, the remaining parameters are chosen such that the cross sections for $ggF\to H \to (H_{S} \to bb) + h$ are relatively large, sometimes just below the upper limits from present constraints from the LHC (and from B-physics and dark matter direct detection), see below.
The cross sections for decays of $h$ such as $h \to \tau\tau$ and $h \to \gamma\gamma$ are closely related to the ones for $h \to bb$ since the branching fractions of $h$ satisfy the combined constraints from ATLAS and CMS \cite{ATLAS:2016neq,CMS:2018uag,ATLAS:2020qdt}. Still, deviations of $\sim 10\%$ from the Standard Model values are possible within these constraints, and sometimes realized within the NMSSM.
Therefore we show in Table~1 the cross sections for $ggF\to H \to (H_{S} \to bb) + (h \to XX)$ separately for $XX = bb, \tau\tau, \gamma\gamma$ for all benchmark points. Points indicated by $^{(1)}$ in the second column in Table~1  (and later in Table~4) have cross sections $ggF\to H\to (H_{S} \to bb) + (h \to \tau\tau)$ at the boundary of the region excluded by CMS in \cite{CMS:2021yci}.
(Points indicated by $^{(2)}$ or $^{(3)}$ have cross sections for resonant Higgs pair production or for $ggF\to A\to Z + (H_{S}\to bb)$ just below the boundary of the region excluded by corresponding searches, see below.)

For illustration we show in Fig.~1 the allowed cross sections for $ggF\to H \to (H_{S} \to bb) + (h \to bb)$ for $M_{H_S}=200$~GeV as function of $M_H$ from Table~1. For $M_H \leq 800$~GeV (red dotted line) these are limited by constraints from $ggF\to A\to Z + (H_{S}\to bb)$ on the parameter space of the NMSSM which explains its irregular shape.

\begin{figure}
\begin{center}
 \includegraphics[scale=1.0]{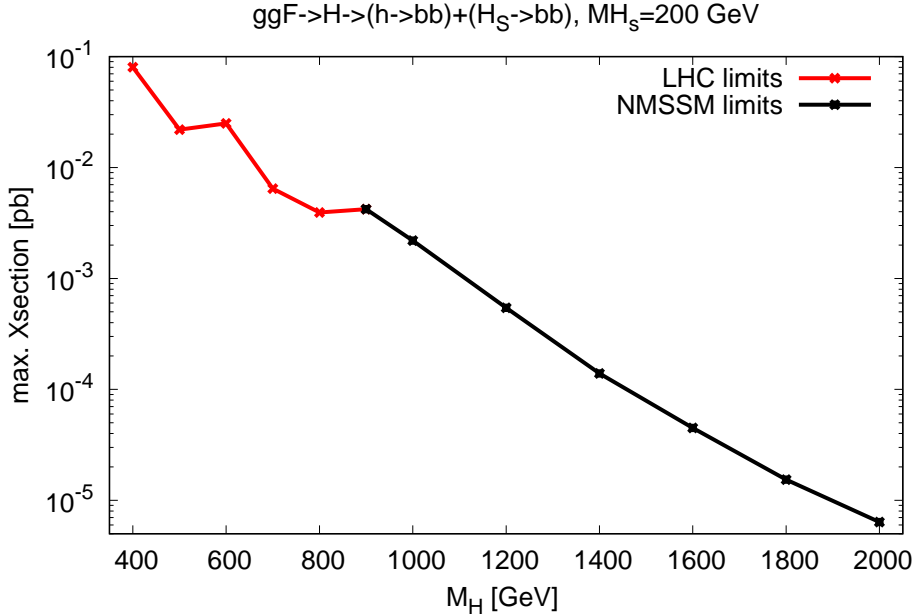}
\end{center}
\caption{Allowed cross sections for $ggF\to H \to (H_{S} \to bb) + (h \to bb)$ for $M_{H_S}=200$~GeV as function of $M_H$. For $M_H \leq 800$~GeV (red dotted line) these are limited by constraints from $ggF\to A\to Z + (H_{S}\to bb)$ \cite{ATLAS:2020gxx}.}
\end{figure}

$H_{S}$ has additional interesting decay modes other than $H_{S} \to bb$. For instance, the branching fraction into $\tau\tau$ is always smaller by a factor $0.1-0.14$, depending on $M_{H_{S}}$. This allows to estimate the cross sections for $H_{S} \to \tau\tau$ from the ones into $bb$.

For $M_{H_{S}}>250$~GeV, the branching fraction $BR(H_{S}\to h+h)$ becomes sizeable (up to $\sim 20\%$), and the cascade $H \to H_{S} + h$ leads to triple Higgs production. 
Furthermore, for $M_{H_{S}}>350$~GeV, the branching fraction $BR(H_{S}\to tt)$ becomes dominant. 
Since both processes are of interest, we added the cross sections for $ggF\to H \to (H_{S} \to h+h) + h$ and for $ggF\to H \to (H_{S} \to tt) + h$ (without branching fractions of $h$ which are within $\sim 10\%$ of the Standard Model values) in Table~1.
The branching fractions for $H_{S} \to h+h$ can become small for accidential cancellations within the corresponding trilinear coupling c) in \eqref{trilin}; of potential interest are the cases where these cross sections are relatively large.

In principle the processes $ggF\to A \to A_S + h$ can lead to identical signatures as the considered processes $ggF\to H \to H_{S} + h$. However, we found in Section~2 that the considered processes have larger cross sections and are thus more promising for potential discoveries (or exclusions). As discussed in the introduction and in Section~2 the final state from $A_S\to \gamma\gamma$ is an exception. In Table~2 we show possible cross sections for $ggF\to A \to (h\to \tau\tau) + (A_S\to \gamma\gamma)$ for $M_A$ near 400, 500, 600 and 700~GeV and $M_{A_S}=70, 100$ and 200~GeV. (For $M_{A_S}=200$~GeV the channel $A_S\to Z+\gamma$ is open, and the branching fraction for $A_s\to \gamma\gamma$ shrinks from $\sim 99\%$ to $\sim 66\%$.)
For illustration we show in Fig.~2 the allowed cross sections for $ggF\to A \to (h\to \tau\tau) + (A_S\to \gamma\gamma)$ for $M_A$ from $\sim 410$ to 700~GeV from Table~2.

\begin{figure}
\begin{center}
 \includegraphics[scale=1.0]{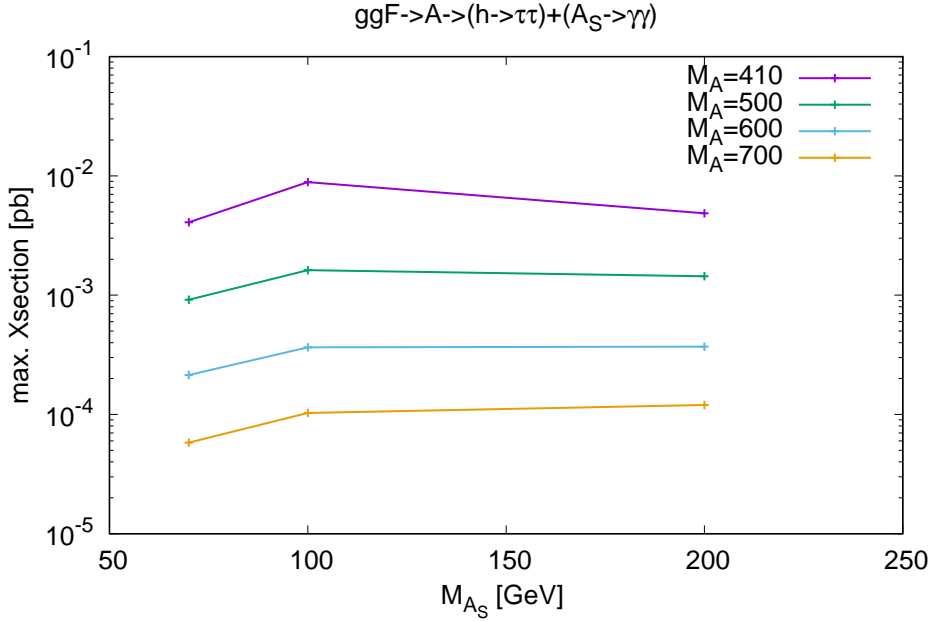}
\end{center}
\caption{Allowed cross sections for $ggF\to A \to (h\to \tau\tau) + (A_S\to \gamma\gamma)$ for $M_A$ from $\sim 410$ to 700~GeV}
\end{figure}

Another interesting process within the NMSSM is resonant (SM) Higgs pair production. The role of the ``resonance'' can be played by $H$ or by $H_{S}$.
The most constraining limits on SM Higgs pair production $ggF\to H\to h+h$ originate from the combination of $bbbb$, $bb\tau\tau$ and $bb\gamma\gamma$ final states by ATLAS in \cite{ATLAS:2021tyg} for $M_H=250-3000$~GeV. 
In fact, for $M_{H}\leq 650$~GeV the cross sections in the NMSSM could be larger than the limits obtained by ATLAS in \cite{ATLAS:2021tyg}, hence these limits constrain the parameter space of the NMSSM. For $M_{H_{S}} \leq 120$~GeV, these limits imply lower bounds on $M_{H}$ (depending on $M_{H_{S}}$) slightly above 400~GeV.
Among the selected benchmark points in Table~1 (and later in Table~4), points indicated by~$^{(2)}$ in the second column (for $M_{H}$ near 400~GeV) have cross sections $ggF\to H\to h + h$ just below the boundary of the region excluded by ATLAS in \cite{ATLAS:2021tyg}.

As discussed in Section~2, particularly large cross sections for resonant SM Higgs pair production can be found if $H$ and $H_{S}$ strongly mix. Then the notation $H_2$, $H_3$ is more appropriate, and the largest cross sections are found for $ggF\to H_2 \to h+h$.
In Table~3 we show possible cross sections in the NMSSM (for points which differ from the benchmark points in Table~1) for $ggF\to H_2 \to h + h$ for $M_{H_2} > 700$~GeV up to $M_{H_2} =1200$~GeV; these are still below the limits from CMS in \cite{CMS:2021roc}.
For illustration we show in Fig.~3 the allowed cross sections for resonant Higgs pair production as function of $M_H$ from Table~3. 

\begin{figure}
\begin{center}
 \includegraphics[scale=1.0]{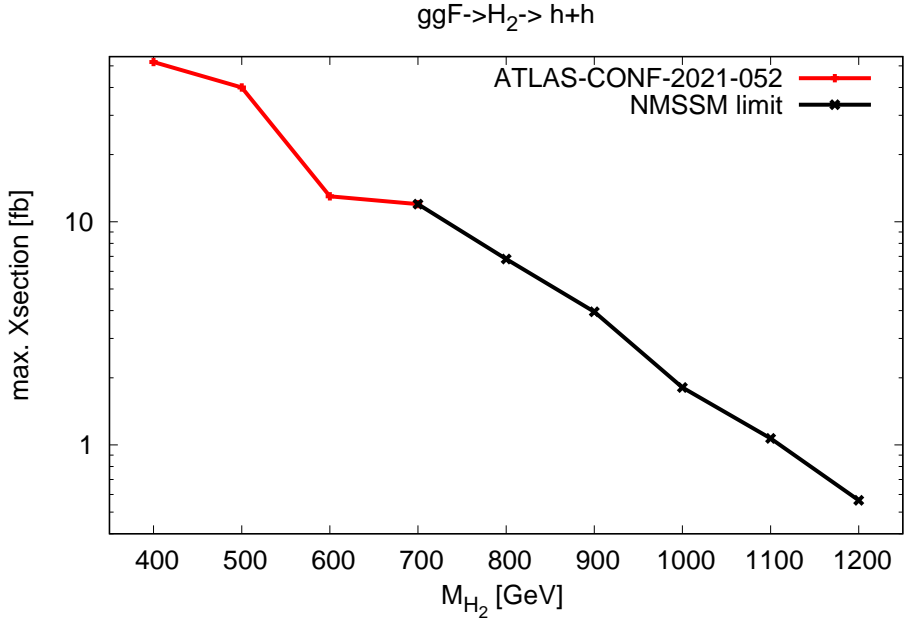}
\end{center}
\caption{Allowed cross sections for resonant Higgs pair production as function of $M_{H_2}$. For $M_{H_2} < 700$~GeV these correspond to the limits from ATLAS in \cite{ATLAS:2021tyg}.}
\end{figure}

Finally we turn to cascade decays into a $Z$ boson. As discussed in Section~2, the largest cross sections in the NMSSM correspond to the processes $ggF\to A \to H_{S} +Z$; cross sections for $ggF\to H \to A_{S} +Z$ are substantially smaller for equivalent masses of $H/A$ and $A_S/H_S$.
Corresponding searches have been performed by CMS in \cite{CMS:2019ogx} after $\sim 36$~fb$^{-1}$ for the mass ranges $120\ \text{GeV}<M_{A}< 1000$~GeV and $30 \ \text{GeV}<M_{H_S}< 780$~GeV, and by ATLAS in \cite{ATLAS:2020gxx} after $\sim 139$~fb$^{-1}$ for $230\ \text{GeV}<M_{A}<  800$~GeV and $130 \ \text{GeV}<M_{H_S}< 700$~GeV. (We have adopted the notation to the interpretation within the NMSSM.)

The latter search excludes some regions in the parameter space of the NMSSM; the benchmark points shown here satisfy these constraints. Points indicated by $^{(3)}$ in the second column of Tables~1 and~4 have cross sections $ggF\to A\to Z + (H_{S}\to bb)$ at the boundary of the region excluded in \cite{ATLAS:2020gxx}. In the Table~4 we show the cross sections for $ggF\to A\to Z + (H_{S}\to XX)$ for various final states $XX=bb,\tau\tau,\gamma\gamma$ for the same benchmark points as in Table~1 which allows to compare the sensitivities in the various search channels. (For some points, the branching fraction for $H_{S}\to \gamma\gamma$ can be particularly small due to cancellations among different loop contributions.)
Also shown in Table~4 are cross sections for the yet unexplored processes $ggF\to A \to Z+(H_{S} \to h+h)$ and $ggF\to A \to Z+(H_{S} \to tt)$ which can possibly be within reach.

Searches for $A \to h+Z$ have been performed by CMS in \cite{CMS:2019qcx,CMS:2019kca} after $\sim 36$~fb$^{-1}$, and by ATLAS in \cite{ATLAS:2020pgp} after $\sim 139$~fb$^{-1}$. In the NMSSM these cross sections are dominated by the production of the MSSM-like pseudo-scalar. In the Table~5 we show the largest possible cross sections in the NMSSM for $400 < M_{A} < 2000$~GeV (for points different from the previous benchmark points) which are well below the present limits obtained by CMS and ATLAS.

\section{Summary}

Searches for Higgs-to-Higgs and Higgs-to-Higgs+Z cascade decays at the LHC allow to explore extended Higgs sectors beyond the SM. In the present paper we have presented various benchmark planes and lines which show which cross sections are possible in which final states within the NMSSM, subject to present phenomenological and theoretical constraints. Some of the available searches by ATLAS and CMS already touch the parameter space of the NMSSM, and our tables allow to estimate which future searches can be promising not only using available data, but also after the upgrade of the LHC to High Luminosity after a suitable rescaling. The proposed search channels $H\to  h+(H_{S} \to t\bar{t})$, $A \to Z+ (H_{S}\to h+h)$ and $A \to Z+ (H_{S}\to t\bar{t})$ are new and have not been considered before.

\section*{Acknowledgements}

U.E. acknowledges motivating and helpful discussions with members of the LHC-HXSWG3-NMSSM working group.


\begin{table}
\centerline{\bf \Large Appendix}

\vspace{3mm}
Tables in different format (csv) as well as SLHA files for the benchmark points are available from the authors upon request.

\begin{center}
\caption{
Possible cross sections (in pb) at 13~TeV for \newline
 $ggF\to H \to (H_{S} \to bb) + (h \to XX)$, $XX = bb, \tau\tau, \gamma\gamma$ (columns 3-5),\newline
 $ggF\to H \to (H_{S} \to h+h) + h$ (column 6),\newline
$ggF\to H \to (H_{S} \to tt) + h$  (column 7).\newline
Points indicated by $^{(1)}$ in the second column have cross sections $ggF\to H\to (H_{S} \to bb) + (h \to \tau\tau)$ at the boundary of the region excluded by CMS in \cite{CMS:2021yci}.\ \newline
Points indicated by $^{(2)}$ in the second column have cross sections $ggF\to H\to h + h$ at the boundary of the region excluded by ATLAS in \cite{ATLAS:2021tyg}.\ \newline
Points indicated by $^{(3)}$ in the second column have cross sections $ggF\to A\to Z + (H_{S}\to bb)$ at the boundary of the region excluded by ATLAS in \cite{ATLAS:2020gxx}.
}

\begin{tabular}{| c | c | c | c | c |c |c | c|}
\hline
$M_{H}$ & $M_{H_{S}}$ & $h\to bb$ & $h\to \tau\tau$ & $h\to \gamma\gamma$ &
$H_{S} \to h + h$ &$H_{S}\to tt$\\
\hline
  408 &   60$^{(2)}$ &  7.112 $\cdot 10^{-2}$ &  7.582 $\cdot 10^{-3}$ &  1.847 $\cdot 10^{-4}$ & - & - \\
  408 &   80$^{(2)}$ &  6.986 $\cdot 10^{-2}$ &  7.447 $\cdot 10^{-3}$ &  1.822 $\cdot 10^{-4}$ & - & - \\
  401 &   90$^{(2)}$ &  1.058 $\cdot 10^{-1}$ &  1.127 $\cdot 10^{-2}$ &  2.805 $\cdot 10^{-4}$ & - & - \\
  401 &  100$^{(2)}$ &  1.060 $\cdot 10^{-1}$ &  1.130 $\cdot 10^{-2}$ &  2.835 $\cdot 10^{-4}$ & - & - \\
  400 &  150$^{(3)}$ &  1.203 $\cdot 10^{-1}$ &  1.282 $\cdot 10^{-2}$ &  3.538 $\cdot 10^{-4}$ & - & - \\
  400 &  200$^{(3)}$ &  8.029 $\cdot 10^{-2}$ &  8.559 $\cdot 10^{-3}$ &  2.182 $\cdot 10^{-4}$ & - & - \\
  400 &  250 &  5.134 $\cdot 10^{-3}$ &  5.473 $\cdot 10^{-4}$ &  1.408 $\cdot 10^{-5}$ & - & - \\
\hline
  500 &   60 &  5.792 $\cdot 10^{-3}$ &  6.172 $\cdot 10^{-4}$ &  1.608 $\cdot 10^{-5}$ & - & - \\
  500 &   80$^{(1)}$ &  3.224 $\cdot 10^{-2}$ &  3.437 $\cdot 10^{-3}$ &  9.136 $\cdot 10^{-5}$ & - & - \\
  500 &   90$^{(1)}$ &  3.727 $\cdot 10^{-2}$ &  3.973 $\cdot 10^{-3}$ &  1.064 $\cdot 10^{-4}$ & - & - \\
  500 &  100 &  1.786 $\cdot 10^{-2}$ &  1.903 $\cdot 10^{-3}$ &  4.555 $\cdot 10^{-5}$ & - & - \\
  500 &  150$^{(3)}$ &  2.222 $\cdot 10^{-2}$ &  2.368 $\cdot 10^{-3}$ &  6.577 $\cdot 10^{-5}$ & - & - \\
  500 &  200$^{(3)}$ &  2.194 $\cdot 10^{-2}$ &  2.341 $\cdot 10^{-3}$ &  7.415 $\cdot 10^{-5}$ & - & - \\
  500 &  250$^{(3)}$ &  4.259 $\cdot 10^{-2}$ &  4.546 $\cdot 10^{-3}$ &  1.487 $\cdot 10^{-4}$ & - & - \\
  500 &  300$^{(3)}$ &  2.617 $\cdot 10^{-2}$ &  2.795 $\cdot 10^{-3}$ &  9.755 $\cdot 10^{-5}$ &  1.570 $\cdot 10^{-3}$ & - \\
\hline
  600 &   60 &  2.913 $\cdot 10^{-3}$ &  3.102 $\cdot 10^{-4}$ &  8.732 $\cdot 10^{-6}$ & - & - \\
  600 &   80$^{(1)}$ &  2.839 $\cdot 10^{-2}$ &  3.026 $\cdot 10^{-3}$ &  9.022 $\cdot 10^{-5}$ & - & - \\
  600 &   90$^{(1)}$ &  2.275 $\cdot 10^{-2}$ &  2.421 $\cdot 10^{-3}$ &  6.950 $\cdot 10^{-5}$ & - & - \\
  600 &  100$^{(1)}$ &  2.237 $\cdot 10^{-2}$ &  2.383 $\cdot 10^{-3}$ &  6.957 $\cdot 10^{-5}$ & - & - \\
  600 &  150$^{(3)}$ &  1.053 $\cdot 10^{-2}$ &  1.122 $\cdot 10^{-3}$ &  3.383 $\cdot 10^{-5}$ & - & - \\
  600 &  200$^{(3)}$ &  2.496 $\cdot 10^{-2}$ &  2.658 $\cdot 10^{-3}$ &  9.316 $\cdot 10^{-5}$ & - & - \\
  600 &  250$^{(3)}$ &  1.185 $\cdot 10^{-2}$ &  1.266 $\cdot 10^{-3}$ &  4.227 $\cdot 10^{-5}$ & - & - \\
  600 &  300$^{(3)}$ &  1.840 $\cdot 10^{-2}$ &  1.965 $\cdot 10^{-3}$ &  7.080 $\cdot 10^{-5}$ &  3.498 $\cdot 10^{-2}$ & - \\
  600 &  400 &  2.707 $\cdot 10^{-3}$ &  2.892 $\cdot 10^{-4}$ &  1.021 $\cdot 10^{-5}$ &  1.564 $\cdot 10^{-3}$ &  9.361 $\cdot 10^{-3}$ \\
\hline
\end{tabular}
\end{center}
\end{table}

\begin{table}[ht]
\caption*{Table 1 continued}
\begin{center}
\begin{tabular}{| c | c | c | c | c |c |c | c|}
\hline
$M_{H}$ & $M_{H_{S}}$ & $h\to bb$ & $h\to \tau\tau$ & $h\to \gamma\gamma$ &
$H_{S} \to h+h$ &$H_{S}\to tt$\\
\hline
  700 &   60 &  1.856 $\cdot 10^{-2}$ &  1.976 $\cdot 10^{-3}$ &  7.059 $\cdot 10^{-5}$ & - & - \\
  700 &   80 &  1.924 $\cdot 10^{-2}$ &  2.048 $\cdot 10^{-3}$ &  7.247 $\cdot 10^{-5}$ & - & - \\
  700 &   90 &  1.941 $\cdot 10^{-2}$ &  2.065 $\cdot 10^{-3}$ &  7.200 $\cdot 10^{-5}$ & - & - \\
  700 &  100 &  1.941 $\cdot 10^{-2}$ &  2.067 $\cdot 10^{-3}$ &  7.276 $\cdot 10^{-5}$ & - & - \\
  700 &  150$^{(3)}$ &  7.745 $\cdot 10^{-3}$ &  8.242 $\cdot 10^{-4}$ &  3.548 $\cdot 10^{-5}$ & - & - \\
  700 &  200$^{(3)}$ &  6.454 $\cdot 10^{-3}$ &  6.880 $\cdot 10^{-4}$ &  2.351 $\cdot 10^{-5}$ & - & - \\
  700 &  250$^{(3)}$ &  1.436 $\cdot 10^{-2}$ &  1.534 $\cdot 10^{-3}$ &  5.573 $\cdot 10^{-5}$ & - & - \\
  700 &  300$^{(3)}$ &  7.742 $\cdot 10^{-3}$ &  8.273 $\cdot 10^{-4}$ &  3.007 $\cdot 10^{-5}$ &  2.319 $\cdot 10^{-2}$ & - \\
  700 &  400 &  3.427 $\cdot 10^{-3}$ &  3.662 $\cdot 10^{-4}$ &  1.287 $\cdot 10^{-5}$ &  3.305 $\cdot 10^{-4}$ &  9.333 $\cdot 10^{-3}$ \\
  700 &  500 &  5.079 $\cdot 10^{-4}$ &  5.430 $\cdot 10^{-5}$ &  1.963 $\cdot 10^{-6}$ &  7.304 $\cdot 10^{-4}$ &  6.373 $\cdot 10^{-3}$ \\
\hline
  800 &   60 &  8.404 $\cdot 10^{-3}$ &  8.961 $\cdot 10^{-4}$ &  3.193 $\cdot 10^{-5}$ & - & - \\
  800 &   80 &  8.447 $\cdot 10^{-3}$ &  9.000 $\cdot 10^{-4}$ &  3.202 $\cdot 10^{-5}$ & - & - \\
  800 &   90 &  8.520 $\cdot 10^{-3}$ &  9.070 $\cdot 10^{-4}$ &  3.167 $\cdot 10^{-5}$ & - & - \\
  800 &  100 &  8.648 $\cdot 10^{-3}$ &  9.207 $\cdot 10^{-4}$ &  3.227 $\cdot 10^{-5}$ & - & - \\
  800 &  150$^{(3)}$ &  2.825 $\cdot 10^{-3}$ &  3.005 $\cdot 10^{-4}$ &  1.332 $\cdot 10^{-5}$ & - & - \\
  800 &  200$^{(3)}$ &  3.934 $\cdot 10^{-3}$ &  4.189 $\cdot 10^{-4}$ &  1.473 $\cdot 10^{-5}$ & - & - \\
  800 &  250$^{(3)}$ &  5.297 $\cdot 10^{-3}$ &  5.636 $\cdot 10^{-4}$ &  1.925 $\cdot 10^{-5}$ & - & - \\
  800 &  300$^{(3)}$ &  3.844 $\cdot 10^{-3}$ &  4.104 $\cdot 10^{-4}$ &  1.479 $\cdot 10^{-5}$ &  1.160 $\cdot 10^{-2}$ & - \\
  800 &  400 &  1.727 $\cdot 10^{-3}$ &  1.839 $\cdot 10^{-4}$ &  6.346 $\cdot 10^{-6}$ &  1.839 $\cdot 10^{-5}$ &  9.374 $\cdot 10^{-3}$ \\
  800 &  500 &  6.218 $\cdot 10^{-4}$ &  6.621 $\cdot 10^{-5}$ &  2.293 $\cdot 10^{-6}$ &  2.574 $\cdot 10^{-4}$ &  8.400 $\cdot 10^{-3}$ \\
  800 &  600 &  1.236 $\cdot 10^{-4}$ &  1.317 $\cdot 10^{-5}$ &  4.618 $\cdot 10^{-7}$ &  1.074 $\cdot 10^{-4}$ &  4.794 $\cdot 10^{-3}$ \\
\hline
  900 &   60 &  4.060 $\cdot 10^{-3}$ &  4.322 $\cdot 10^{-4}$ &  1.466 $\cdot 10^{-5}$ & - & - \\
  900 &   80 &  4.081 $\cdot 10^{-3}$ &  4.341 $\cdot 10^{-4}$ &  1.503 $\cdot 10^{-5}$ & - & - \\
  900 &   90 &  4.108 $\cdot 10^{-3}$ &  4.370 $\cdot 10^{-4}$ &  1.506 $\cdot 10^{-5}$ & - & - \\
  900 &  100 &  4.056 $\cdot 10^{-3}$ &  4.315 $\cdot 10^{-4}$ &  1.489 $\cdot 10^{-5}$ & - & - \\
  900 &  150 &  4.203 $\cdot 10^{-3}$ &  4.472 $\cdot 10^{-4}$ &  1.513 $\cdot 10^{-5}$ & - & - \\
  900 &  200 &  4.209 $\cdot 10^{-3}$ &  4.479 $\cdot 10^{-4}$ &  1.499 $\cdot 10^{-5}$ & - & - \\
  900 &  250 &  3.157 $\cdot 10^{-3}$ &  3.377 $\cdot 10^{-4}$ &  1.288 $\cdot 10^{-5}$ & - & - \\
  900 &  300 &  3.390 $\cdot 10^{-3}$ &  3.606 $\cdot 10^{-4}$ &  1.253 $\cdot 10^{-5}$ &  3.586 $\cdot 10^{-6}$ & - \\
  900 &  400 &  5.960 $\cdot 10^{-4}$ &  6.342 $\cdot 10^{-5}$ &  2.156 $\cdot 10^{-6}$ &  5.370 $\cdot 10^{-6}$ &  3.933 $\cdot 10^{-3}$ \\
  900 &  500 &  3.302 $\cdot 10^{-4}$ &  3.512 $\cdot 10^{-5}$ &  1.152 $\cdot 10^{-6}$ &  3.281 $\cdot 10^{-7}$ &  3.232 $\cdot 10^{-3}$ \\
  900 &  600 &  1.803 $\cdot 10^{-4}$ &  1.919 $\cdot 10^{-5}$ &  6.271 $\cdot 10^{-7}$ &  1.790 $\cdot 10^{-5}$ &  2.556 $\cdot 10^{-3}$ \\
900 &  700 &  2.469 $\cdot 10^{-5}$ &  2.639 $\cdot 10^{-6}$ &  9.232 $\cdot 10^{-8}$ &  1.209 $\cdot 10^{-5}$ &   5.133 $\cdot 10^{-4}$ \\
\hline
\end{tabular}
\end{center}
\end{table}

\begin{table}[ht]
\caption*{Table 1 continued}
\begin{center}
\begin{tabular}{| c | c | c | c | c |c |c | c|}
\hline
$M_{H}$ & $M_{H_{S}}$ & $h\to bb$ & $h\to \tau\tau$ & $h\to \gamma\gamma$ &
$H_{S} \to h+h$ &$H_{S}\to tt$\\
\hline
 1000 &   60 &  1.893 $\cdot 10^{-3}$ &  2.014 $\cdot 10^{-4}$ &  6.774 $\cdot 10^{-6}$ & - & - \\
 1000 &   80 &  1.905 $\cdot 10^{-3}$ &  2.027 $\cdot 10^{-4}$ &  6.978 $\cdot 10^{-6}$ & - & - \\
 1000 &   90 &  1.917 $\cdot 10^{-3}$ &  2.039 $\cdot 10^{-4}$ &  7.022 $\cdot 10^{-6}$ & - & - \\
 1000 &  100 &  1.888 $\cdot 10^{-3}$ &  2.008 $\cdot 10^{-4}$ &  7.057 $\cdot 10^{-6}$ & - & - \\
 1000 &  150 &  2.062 $\cdot 10^{-3}$ &  2.194 $\cdot 10^{-4}$ &  7.513 $\cdot 10^{-6}$ & - & - \\
 1000 &  200 &  2.196 $\cdot 10^{-3}$ &  2.338 $\cdot 10^{-4}$ &  7.634 $\cdot 10^{-6}$ & - & - \\
 1000 &  250 &  1.442 $\cdot 10^{-3}$ &  1.537 $\cdot 10^{-4}$ &  5.518 $\cdot 10^{-6}$ & - & - \\
 1000 &  300 &  1.366 $\cdot 10^{-3}$ &  1.453 $\cdot 10^{-4}$ &  5.045 $\cdot 10^{-6}$ &  3.186 $\cdot 10^{-4}$ & - \\
 1000 &  400 &  3.429 $\cdot 10^{-4}$ &  3.648 $\cdot 10^{-5}$ &  1.247 $\cdot 10^{-6}$ &  3.837 $\cdot 10^{-6}$ &  2.084 $\cdot 10^{-3}$ \\
 1000 &  500 &  1.725 $\cdot 10^{-4}$ &  1.836 $\cdot 10^{-5}$ &  6.153 $\cdot 10^{-7}$ &  9.341 $\cdot 10^{-6}$ &  2.184 $\cdot 10^{-3}$ \\
 1000 &  600 &  1.328 $\cdot 10^{-4}$ &  1.413 $\cdot 10^{-5}$ &  4.706 $\cdot 10^{-7}$ &  7.819 $\cdot 10^{-6}$ &  1.694 $\cdot 10^{-3}$ \\
 1000 &  700 &  6.114 $\cdot 10^{-5}$ &  6.509 $\cdot 10^{-6}$ &  2.195 $\cdot 10^{-7}$ &  1.736 $\cdot 10^{-4}$ &  9.241 $\cdot 10^{-4}$ \\
 1000 &  800 &  6.678 $\cdot 10^{-6}$ &  7.106 $\cdot 10^{-7}$ &  2.355 $\cdot 10^{-8}$ &  1.629 $\cdot 10^{-7}$ &  2.034 $\cdot 10^{-4}$ \\
\hline
 1200 &   60 &  5.087 $\cdot 10^{-4}$ &  5.418 $\cdot 10^{-5}$ &  1.834 $\cdot 10^{-6}$ & - & - \\
 1200 &   80 &  5.125 $\cdot 10^{-4}$ &  5.456 $\cdot 10^{-5}$ &  1.895 $\cdot 10^{-6}$ & - & - \\
 1200 &  100 &  5.033 $\cdot 10^{-4}$ &  5.353 $\cdot 10^{-5}$ &  1.833 $\cdot 10^{-6}$ & - & - \\
 1200 &  150 &  4.917 $\cdot 10^{-4}$ &  5.263 $\cdot 10^{-5}$ &  1.933 $\cdot 10^{-6}$ & - & - \\
 1200 &  200 &  5.434 $\cdot 10^{-4}$ &  5.783 $\cdot 10^{-5}$ &  1.902 $\cdot 10^{-6}$ & - & - \\
 1200 &  300 &  4.484 $\cdot 10^{-4}$ &  4.770 $\cdot 10^{-5}$ &  1.632 $\cdot 10^{-6}$ &  8.027 $\cdot 10^{-5}$ & - \\
 1200 &  400 &  1.108 $\cdot 10^{-4}$ &  1.178 $\cdot 10^{-5}$ &  3.999 $\cdot 10^{-7}$ &  2.211 $\cdot 10^{-5}$ &  5.664 $\cdot 10^{-4}$ \\
 1200 &  500 &  5.677 $\cdot 10^{-5}$ &  6.040 $\cdot 10^{-6}$ &  2.025 $\cdot 10^{-7}$ &  1.249 $\cdot 10^{-5}$ &  6.287 $\cdot 10^{-4}$ \\
 1200 &  600 &  4.245 $\cdot 10^{-5}$ &  4.518 $\cdot 10^{-6}$ &  1.497 $\cdot 10^{-7}$ &  1.769 $\cdot 10^{-5}$ &  6.805 $\cdot 10^{-4}$ \\
 1200 &  700 &  3.157 $\cdot 10^{-5}$ &  3.360 $\cdot 10^{-6}$ &  1.122 $\cdot 10^{-7}$ &  8.340 $\cdot 10^{-7}$ &  6.422 $\cdot 10^{-4}$ \\
 1200 &  800 &  2.214 $\cdot 10^{-5}$ &  2.356 $\cdot 10^{-6}$ &  7.851 $\cdot 10^{-8}$ &  4.322 $\cdot 10^{-6}$ &  6.238 $\cdot 10^{-4}$ \\
\hline
 1400 &   60 &  1.595 $\cdot 10^{-4}$ &  1.702 $\cdot 10^{-5}$ &  5.937 $\cdot 10^{-7}$ & - & - \\
 1400 &   80 &  1.677 $\cdot 10^{-4}$ &  1.786 $\cdot 10^{-5}$ &  6.258 $\cdot 10^{-7}$ & - & - \\
 1400 &  100 &  1.543 $\cdot 10^{-4}$ &  1.646 $\cdot 10^{-5}$ &  5.871 $\cdot 10^{-7}$ & - & - \\
 1400 &  150 &  1.562 $\cdot 10^{-4}$ &  1.662 $\cdot 10^{-5}$ &  5.650 $\cdot 10^{-7}$ & - & - \\
 1400 &  200 &  1.392 $\cdot 10^{-4}$ &  1.481 $\cdot 10^{-5}$ &  4.901 $\cdot 10^{-7}$ & - & - \\
 1400 &  300 &  1.295 $\cdot 10^{-4}$ &  1.378 $\cdot 10^{-5}$ &  4.716 $\cdot 10^{-7}$ &  3.941 $\cdot 10^{-5}$ & - \\
 1400 &  400 &  4.102 $\cdot 10^{-5}$ &  4.364 $\cdot 10^{-6}$ &  1.493 $\cdot 10^{-7}$ &  1.035 $\cdot 10^{-5}$ &  1.598 $\cdot 10^{-4}$ \\
 1400 &  500 &  2.150 $\cdot 10^{-5}$ &  2.288 $\cdot 10^{-6}$ &  7.743 $\cdot 10^{-8}$ &  5.671 $\cdot 10^{-6}$ &  1.921 $\cdot 10^{-4}$ \\
 1400 &  600 &  1.619 $\cdot 10^{-5}$ &  1.722 $\cdot 10^{-6}$ &  5.779 $\cdot 10^{-8}$ &  9.012 $\cdot 10^{-6}$ &  2.207 $\cdot 10^{-4}$ \\
 1400 &  700 &  9.399 $\cdot 10^{-6}$ &  1.000 $\cdot 10^{-6}$ &  3.386 $\cdot 10^{-8}$ &  2.049 $\cdot 10^{-5}$ &  8.678 $\cdot 10^{-5}$ \\
 1400 &  800 &  4.745 $\cdot 10^{-6}$ &  5.079 $\cdot 10^{-7}$ &  1.879 $\cdot 10^{-8}$ &  3.354 $\cdot 10^{-5}$ &  3.992 $\cdot 10^{-5}$ \\
\hline
\end{tabular}
\end{center}
\end{table}

\begin{table}[ht]
\caption*{Table 1 continued}
\begin{center}
\begin{tabular}{| c | c | c | c | c |c |c | c|}
\hline
$M_{H}$ & $M_{H_{S}}$ & $h\to bb$ & $h\to \tau\tau$ & $h\to \gamma\gamma$ &
$H_{S} \to h+h$ &$H_{S}\to tt$\\
\hline
 1600 &   60 &  4.685 $\cdot 10^{-5}$ &  4.995 $\cdot 10^{-6}$ &  1.710 $\cdot 10^{-7}$ & - & - \\
 1600 &   80 &  4.723 $\cdot 10^{-5}$ &  5.035 $\cdot 10^{-6}$ &  1.809 $\cdot 10^{-7}$ & - & - \\
 1600 &  100 &  4.387 $\cdot 10^{-5}$ &  4.673 $\cdot 10^{-6}$ &  1.756 $\cdot 10^{-7}$ & - & - \\
 1600 &  150 &  4.293 $\cdot 10^{-5}$ &  4.566 $\cdot 10^{-6}$ &  1.567 $\cdot 10^{-7}$ & - & - \\
 1600 &  200 &  4.493 $\cdot 10^{-5}$ &  4.781 $\cdot 10^{-6}$ &  1.595 $\cdot 10^{-7}$ & - & - \\
 1600 &  300 &  3.124 $\cdot 10^{-5}$ &  3.323 $\cdot 10^{-6}$ &  1.138 $\cdot 10^{-7}$ &  1.537 $\cdot 10^{-5}$ & - \\
 1600 &  400 &  1.261 $\cdot 10^{-5}$ &  1.342 $\cdot 10^{-6}$ &  4.633 $\cdot 10^{-8}$ &  3.343 $\cdot 10^{-6}$ &  3.869 $\cdot 10^{-5}$ \\
 1600 &  500 &  6.427 $\cdot 10^{-6}$ &  6.837 $\cdot 10^{-7}$ &  2.332 $\cdot 10^{-8}$ &  2.602 $\cdot 10^{-6}$ &  4.814 $\cdot 10^{-5}$ \\
 1600 &  600 &  4.705 $\cdot 10^{-6}$ &  5.005 $\cdot 10^{-7}$ &  1.690 $\cdot 10^{-8}$ &  5.436 $\cdot 10^{-6}$ &  5.417 $\cdot 10^{-5}$ \\
 1600 &  700 &  1.827 $\cdot 10^{-6}$ &  1.949 $\cdot 10^{-7}$ &  6.878 $\cdot 10^{-9}$ &  4.369 $\cdot 10^{-6}$ &  7.397 $\cdot 10^{-6}$ \\
 1600 &  800 &  1.097 $\cdot 10^{-6}$ &  1.170 $\cdot 10^{-7}$ &  4.077 $\cdot 10^{-9}$ &  1.117 $\cdot 10^{-5}$ &  3.752 $\cdot 10^{-6}$ \\
\hline
 1800 &   60 &  1.646 $\cdot 10^{-5}$ &  1.752 $\cdot 10^{-6}$ &  5.836 $\cdot 10^{-8}$ & - & - \\
 1800 &   80 &  1.681 $\cdot 10^{-5}$ &  1.788 $\cdot 10^{-6}$ &  6.117 $\cdot 10^{-8}$ & - & - \\
 1800 &  100 &  1.650 $\cdot 10^{-5}$ &  1.757 $\cdot 10^{-6}$ &  6.081 $\cdot 10^{-8}$ & - & - \\
 1800 &  150 &  1.494 $\cdot 10^{-5}$ &  1.589 $\cdot 10^{-6}$ &  5.463 $\cdot 10^{-8}$ & - & - \\
 1800 &  200 &  1.537 $\cdot 10^{-5}$ &  1.636 $\cdot 10^{-6}$ &  5.534 $\cdot 10^{-8}$ & - & - \\
 1800 &  300 &  8.187 $\cdot 10^{-6}$ &  8.708 $\cdot 10^{-7}$ &  2.982 $\cdot 10^{-8}$ &  7.933 $\cdot 10^{-6}$ & - \\
 1800 &  400 &  4.731 $\cdot 10^{-6}$ &  5.033 $\cdot 10^{-7}$ &  1.723 $\cdot 10^{-8}$ &  1.246 $\cdot 10^{-6}$ &  1.200 $\cdot 10^{-5}$ \\
 1800 &  500 &  2.567 $\cdot 10^{-6}$ &  2.730 $\cdot 10^{-7}$ &  9.341 $\cdot 10^{-9}$ &  1.156 $\cdot 10^{-7}$ &  1.666 $\cdot 10^{-5}$ \\
 1800 &  600 &  1.898 $\cdot 10^{-6}$ &  2.019 $\cdot 10^{-7}$ &  6.901 $\cdot 10^{-9}$ &  2.693 $\cdot 10^{-8}$ &  1.880 $\cdot 10^{-5}$ \\
 1800 &  700 &  4.127 $\cdot 10^{-7}$ &  4.390 $\cdot 10^{-8}$ &  1.510 $\cdot 10^{-9}$ &  5.860 $\cdot 10^{-6}$ &  5.633 $\cdot 10^{-7}$ \\
 1800 &  800 &  2.237 $\cdot 10^{-7}$ &  2.379 $\cdot 10^{-8}$ &  8.196 $\cdot 10^{-10}$ &  7.165 $\cdot 10^{-6}$ &  1.242 $\cdot 10^{-7}$ \\
\hline
 2000 &   60 &  6.318 $\cdot 10^{-6}$ &  6.722 $\cdot 10^{-7}$ &  2.232 $\cdot 10^{-8}$ & - & - \\
 2000 &   80 &  6.409 $\cdot 10^{-6}$ &  6.817 $\cdot 10^{-7}$ &  2.403 $\cdot 10^{-8}$ & - & - \\
 2000 &  100 &  6.335 $\cdot 10^{-6}$ &  6.737 $\cdot 10^{-7}$ &  2.394 $\cdot 10^{-8}$ & - & - \\
 2000 &  150 &  6.583 $\cdot 10^{-6}$ &  7.003 $\cdot 10^{-7}$ &  2.388 $\cdot 10^{-8}$ & - & - \\
 2000 &  200 &  6.366 $\cdot 10^{-6}$ &  6.774 $\cdot 10^{-7}$ &  2.314 $\cdot 10^{-8}$ & - & - \\
 2000 &  300 &  1.430 $\cdot 10^{-6}$ &  1.521 $\cdot 10^{-7}$ &  5.217 $\cdot 10^{-9}$ &  5.406 $\cdot 10^{-6}$ & - \\
 2000 &  400 &  1.927 $\cdot 10^{-6}$ &  2.050 $\cdot 10^{-7}$ &  7.025 $\cdot 10^{-9}$ &  7.781 $\cdot 10^{-7}$ &  3.990 $\cdot 10^{-6}$ \\
 2000 &  500 &  1.242 $\cdot 10^{-6}$ &  1.321 $\cdot 10^{-7}$ &  4.529 $\cdot 10^{-9}$ &  3.862 $\cdot 10^{-10}$ &  7.134 $\cdot 10^{-6}$ \\
 2000 &  600 &  1.076 $\cdot 10^{-6}$ &  1.145 $\cdot 10^{-7}$ &  3.935 $\cdot 10^{-9}$ &  2.395 $\cdot 10^{-7}$ &  9.240 $\cdot 10^{-6}$ \\
 2000 &  700 &  6.211 $\cdot 10^{-7}$ &  6.616 $\cdot 10^{-8}$ &  2.341 $\cdot 10^{-9}$ &  2.359 $\cdot 10^{-6}$ &  4.109 $\cdot 10^{-6}$ \\
 2000 &  800 &  1.730 $\cdot 10^{-7}$ &  1.841 $\cdot 10^{-8}$ &  6.394 $\cdot 10^{-10}$ &  4.016 $\cdot 10^{-6}$ &  2.455 $\cdot 10^{-7}$ \\
\hline
\end{tabular}
\end{center}

\end{table}


\begin{table}[ht]
\caption{Possible cross sections $\sigma$ at 13~TeV for $ggF\to A \to (h\to \tau\tau) + (A_S\to \gamma\gamma)$}
\begin{center}
\begin{tabular}{| c | c | c | }
\hline
$M_{A}$ [GeV] & $M_{A_S}$ [GeV]  & $\sigma$ [fb]  \\
\hline
410 & 70 & 4.08  \\
405 & 100 & 8.85  \\
413 & 200 & 4.06  \\
500 & 70 & 0.916  \\
500 & 100 & 1.62  \\
500 & 200 & 1.26  \\
600 & 70 & 0.214  \\
600 & 100 & 0.365  \\
600 & 200 & 0.370  \\
700 & 70 & 0.0580  \\
700 & 100 & 0.103  \\
700 & 200 & 0.120  \\
\hline
\end{tabular}
\end{center}
\end{table}

\begin{table}
\caption{Possible cross sections $\sigma$ at 13~TeV for $ggF\to H_2 \to h + h$}
\begin{center}
\begin{tabular}{| c | c | c | c | c |c |c | c|}
\hline
$M_{H_2}$ [GeV] & 700 & 800 & 900 & 1000 & 1100 & 1200 \\
\hline
$\sigma$ [fb] & 12.0 & 6.81 & 3.95 & 1.81 & 1.07 & 0.565 \\
\hline
\end{tabular}
\end{center}
\end{table}

\newpage

\begin{table}[ht]
\begin{center}
\caption{Possible cross sections (in pb) at 13~TeV for \newline
 $ggF\to A \to Z + (H_{S} \to XX)$, $XX = bb, \tau\tau, \gamma\gamma$ (columns 3-5)\newline
 $ggF\to A \to Z+ (H_{S} \to h+h)$ (column 6)\newline
$ggF\to A \to Z+(H_{S} \to tt)$  (column 7)\newline
Points indicated by $^{(1)}$ in the second column have cross sections $ggF\to H_3\to (H_{S} \to bb) + (h \to \tau\tau)$ at the boundary of the region excluded by CMS in \cite{CMS:2021yci}.\ \newline
Points indicated by $^{(2)}$ in the second column have cross sections $ggF\to H_3\to h + h$ at the boundary of the region excluded by ATLAS in \cite{ATLAS:2021tyg}.\ \newline
Points indicated by $^{(3)}$ in the second column have cross sections $ggF\to A_2\to Z + (H_{S}\to bb)$ at the boundary of the region excluded by ATLAS in \cite{ATLAS:2020gxx}.
}

\begin{tabular}{| c | c | c | c | c |c |c | c|}
\hline
$M_{A}$ & $M_{H_{S}}$ & $H_{S}\to bb$ & $H_{S}\to \tau\tau$ & $H_{S}\to \gamma\gamma$ &
$H_{S} \to h + h$ &$H_{S}\to tt$\\
\hline
  404 &   60$^{(2)}$ &  1.129 $\cdot 10^{-1}$ &  1.052 $\cdot 10^{-2}$ &  6.083 $\cdot 10^{-6}$ &  - &  - \\
  405 &   80$^{(2)}$ &  1.097 $\cdot 10^{-1}$ &  1.082 $\cdot 10^{-2}$ &  9.590 $\cdot 10^{-6}$ &  - &  - \\
  396 &   90$^{(2)}$ &  1.751 $\cdot 10^{-1}$ &  1.768 $\cdot 10^{-2}$ &  1.201 $\cdot 10^{-5}$ &  - &  - \\
  397 &  100$^{(2)}$ &  1.751 $\cdot 10^{-1}$ &  1.806 $\cdot 10^{-2}$ &  1.179 $\cdot 10^{-5}$ &  - &  - \\
  391 &  150$^{(3)}$ &  1.289 $\cdot 10^{-1}$ &  1.437 $\cdot 10^{-2}$ &  6.428 $\cdot 10^{-7}$ &  - &  - \\
  392 &  200$^{(3)}$ &  1.072 $\cdot 10^{-1}$ &  1.262 $\cdot 10^{-2}$ &  2.038 $\cdot 10^{-5}$ &  - &  - \\
  391 &  250 &  6.626 $\cdot 10^{-3}$ &  8.115 $\cdot 10^{-4}$ &  1.539 $\cdot 10^{-6}$ &  9.927 $\cdot 10^{-4}$ &  - \\
\hline
  497 &   60 &  1.241 $\cdot 10^{-2}$ &  1.171 $\cdot 10^{-3}$ &  2.628 $\cdot 10^{-5}$ &  - &  - \\
  495 &   80$^{(1)}$ &  8.329 $\cdot 10^{-2}$ &  8.402 $\cdot 10^{-3}$ &  1.551 $\cdot 10^{-3}$ &  - &  - \\
  495 &   90$^{(1)}$ &  9.247 $\cdot 10^{-2}$ &  9.519 $\cdot 10^{-3}$ &  1.624 $\cdot 10^{-3}$ &  - &  - \\
  496 &  100 &  3.658 $\cdot 10^{-2}$ &  3.810 $\cdot 10^{-3}$ &  1.468 $\cdot 10^{-4}$ &  - &  - \\
  494 &  150$^{(3)}$ &  4.676 $\cdot 10^{-2}$ &  5.262 $\cdot 10^{-3}$ &  5.796 $\cdot 10^{-4}$ &  - &  - \\
  492 &  200$^{(3)}$ &  4.249 $\cdot 10^{-2}$ &  5.005 $\cdot 10^{-3}$ &  5.573 $\cdot 10^{-5}$ &  - &  - \\
  493 &  250$^{(3)}$ &  8.519 $\cdot 10^{-2}$ &  1.043 $\cdot 10^{-2}$ &  4.684 $\cdot 10^{-5}$ &  7.458 $\cdot 10^{-2}$ &  - \\
  499 &  300$^{(3)}$ &  5.078 $\cdot 10^{-2}$ &  6.420 $\cdot 10^{-3}$ &  1.292 $\cdot 10^{-5}$ &  1.847 $\cdot 10^{-3}$ &  - \\
\hline
  598 &   60 &  6.181 $\cdot 10^{-3}$ &  5.811 $\cdot 10^{-4}$ &  8.353 $\cdot 10^{-6}$ &  - &  - \\
  596 &   80$^{(1)}$ &  6.387 $\cdot 10^{-2}$ &  6.344 $\cdot 10^{-3}$ &  2.545 $\cdot 10^{-4}$ &  - &  - \\
  596 &   90$^{(1)}$ &  5.204 $\cdot 10^{-2}$ &  5.318 $\cdot 10^{-3}$ &  4.580 $\cdot 10^{-4}$ &  - &  - \\
  596 &  100$^{(1)}$ &  5.168 $\cdot 10^{-2}$ &  5.392 $\cdot 10^{-3}$ &  5.399 $\cdot 10^{-4}$ &  - &  - \\
  595 &  150$^{(3)}$ &  2.193 $\cdot 10^{-2}$ &  2.464 $\cdot 10^{-3}$ &  2.390 $\cdot 10^{-4}$ &  - &  - \\
  593 &  200$^{(3)}$ &  5.064 $\cdot 10^{-2}$ &  5.953 $\cdot 10^{-3}$ &  1.083 $\cdot 10^{-5}$ &  - &  - \\
  594 &  250$^{(3)}$ &  2.396 $\cdot 10^{-2}$ &  2.931 $\cdot 10^{-3}$ &  2.360 $\cdot 10^{-5}$ &  - &  - \\
  592 &  300$^{(3)}$ &  4.428 $\cdot 10^{-2}$ &  5.602 $\cdot 10^{-3}$ &  2.690 $\cdot 10^{-5}$ &  5.003 $\cdot 10^{-2}$ &  - \\
  590 &  400 &  1.042 $\cdot 10^{-2}$ &  1.383 $\cdot 10^{-3}$ &  3.142 $\cdot 10^{-6}$ &  3.627 $\cdot 10^{-3}$ &  2.170 $\cdot 10^{-2}$ \\
\hline
\end{tabular}
\end{center}
\end{table}

\begin{table}[ht]
\caption*{Table 4 continued}
\begin{center}
\begin{tabular}{| c | c | c | c | c |c |c | c|}
\hline
$M_{A}$ & $M_{H_{S}}$ & $H_{S}\to bb$ & $H_{S}\to \tau\tau$ & $H_{S}\to \gamma\gamma$ &
$H_{S} \to h + h$ &$H_{S}\to tt$\\
\hline
  697 &   60 &  3.567 $\cdot 10^{-2}$ &  3.308 $\cdot 10^{-3}$ &  4.992 $\cdot 10^{-7}$ &  - &  - \\
  696 &   80 &  3.661 $\cdot 10^{-2}$ &  3.604 $\cdot 10^{-3}$ &  7.707 $\cdot 10^{-7}$ &  - &  - \\
  696 &   90 &  3.668 $\cdot 10^{-2}$ &  3.699 $\cdot 10^{-3}$ &  1.106 $\cdot 10^{-6}$ &  - &  - \\
  696 &  100 &  3.699 $\cdot 10^{-2}$ &  3.808 $\cdot 10^{-3}$ &  1.433 $\cdot 10^{-6}$ &  - &  - \\
  696 &  150$^{(3)}$ &  1.871 $\cdot 10^{-2}$ &  2.078 $\cdot 10^{-3}$ &  2.174 $\cdot 10^{-5}$ &  - &  - \\
  695 &  200$^{(3)}$ &  1.270 $\cdot 10^{-2}$ &  1.495 $\cdot 10^{-3}$ &  1.441 $\cdot 10^{-5}$ &  - &  - \\
  693 &  250$^{(3)}$ &  3.058 $\cdot 10^{-2}$ &  3.744 $\cdot 10^{-3}$ &  2.278 $\cdot 10^{-5}$ &  - &  - \\
  692 &  300$^{(3)}$ &  1.749 $\cdot 10^{-2}$ &  2.211 $\cdot 10^{-3}$ &  1.252 $\cdot 10^{-5}$ &  3.087 $\cdot 10^{-2}$ &  - \\
  689 &  400 &  9.329 $\cdot 10^{-3}$ &  1.237 $\cdot 10^{-3}$ &  6.345 $\cdot 10^{-6}$ &  5.393 $\cdot 10^{-4}$ &  1.523 $\cdot 10^{-2}$ \\
  689 &  500 &  2.059 $\cdot 10^{-3}$ &  2.834 $\cdot 10^{-4}$ &  9.098 $\cdot 10^{-7}$ &  1.748 $\cdot 10^{-3}$ &  1.525 $\cdot 10^{-2}$ \\
\hline
  797 &   60 &  1.633 $\cdot 10^{-2}$ &  1.516 $\cdot 10^{-3}$ &  4.556 $\cdot 10^{-7}$ &  - &  - \\
  797 &   80 &  1.633 $\cdot 10^{-2}$ &  1.608 $\cdot 10^{-3}$ &  4.850 $\cdot 10^{-7}$ &  - &  - \\
  797 &   90 &  1.631 $\cdot 10^{-2}$ &  1.645 $\cdot 10^{-3}$ &  7.845 $\cdot 10^{-7}$ &  - &  - \\
  796 &  100 &  1.661 $\cdot 10^{-2}$ &  1.709 $\cdot 10^{-3}$ &  8.285 $\cdot 10^{-7}$ &  - &  - \\
  797 &  150$^{(3)}$ &  7.514 $\cdot 10^{-3}$ &  8.342 $\cdot 10^{-4}$ &  1.136 $\cdot 10^{-5}$ &  - &  - \\
  796 &  200$^{(3)}$ &  7.853 $\cdot 10^{-3}$ &  9.227 $\cdot 10^{-4}$ &  1.550 $\cdot 10^{-6}$ &  - &  - \\
  795 &  250$^{(3)}$ &  1.043 $\cdot 10^{-2}$ &  1.276 $\cdot 10^{-3}$ &  6.766 $\cdot 10^{-6}$ &  9.927 $\cdot 10^{-3}$ &  - \\
  794 &  300$^{(3)}$ &  8.016 $\cdot 10^{-3}$ &  1.012 $\cdot 10^{-3}$ &  6.540 $\cdot 10^{-6}$ &  1.444 $\cdot 10^{-2}$ &  - \\
  790 &  400 &  4.061 $\cdot 10^{-3}$ &  5.387 $\cdot 10^{-4}$ &  3.293 $\cdot 10^{-6}$ &  2.694 $\cdot 10^{-5}$ &  1.373 $\cdot 10^{-2}$ \\
  788 &  500 &  1.813 $\cdot 10^{-3}$ &  2.498 $\cdot 10^{-4}$ &  1.802 $\cdot 10^{-6}$ &  4.671 $\cdot 10^{-4}$ &  1.524 $\cdot 10^{-2}$ \\
  789 &  600 &  4.709 $\cdot 10^{-4}$ &  6.693 $\cdot 10^{-5}$ &  3.398 $\cdot 10^{-7}$ &  2.509 $\cdot 10^{-4}$ &  1.120 $\cdot 10^{-2}$ \\
\hline
  897 &   60 &  7.742 $\cdot 10^{-3}$ &  7.191 $\cdot 10^{-4}$ &  4.836 $\cdot 10^{-7}$ &  - &  - \\
  897 &   80 &  7.762 $\cdot 10^{-3}$ &  7.633 $\cdot 10^{-4}$ &  2.260 $\cdot 10^{-7}$ &  - &  - \\
  897 &   90 &  7.810 $\cdot 10^{-3}$ &  7.882 $\cdot 10^{-4}$ &  3.683 $\cdot 10^{-7}$ &  - &  - \\
  897 &  100 &  7.750 $\cdot 10^{-3}$ &  7.986 $\cdot 10^{-4}$ &  5.046 $\cdot 10^{-7}$ &  - &  - \\
  896 &  150 &  7.890 $\cdot 10^{-3}$ &  8.798 $\cdot 10^{-4}$ &  1.803 $\cdot 10^{-6}$ &  - &  - \\
  896 &  200 &  7.846 $\cdot 10^{-3}$ &  9.223 $\cdot 10^{-4}$ &  3.948 $\cdot 10^{-6}$ &  - &  - \\
  894 &  250 &  7.507 $\cdot 10^{-3}$ &  9.172 $\cdot 10^{-4}$ &  4.594 $\cdot 10^{-6}$ &  - &  - \\
  894 &  300 &  7.404 $\cdot 10^{-3}$ &  9.325 $\cdot 10^{-4}$ &  5.908 $\cdot 10^{-6}$ &  4.972 $\cdot 10^{-6}$ &  - \\
  894 &  400 &  1.258 $\cdot 10^{-3}$ &  1.664 $\cdot 10^{-4}$ &  1.215 $\cdot 10^{-6}$ &  7.156 $\cdot 10^{-6}$ &  5.241 $\cdot 10^{-3}$ \\
  893 &  500 &  6.878 $\cdot 10^{-4}$ &  9.440 $\cdot 10^{-5}$ &  9.355 $\cdot 10^{-7}$ &  4.382 $\cdot 10^{-7}$ &  4.317 $\cdot 10^{-3}$ \\
  894 &  600 &  3.955 $\cdot 10^{-4}$ &  5.592 $\cdot 10^{-5}$ &  5.944 $\cdot 10^{-7}$ &  2.509 $\cdot 10^{-5}$ &  3.583 $\cdot 10^{-3}$ \\
  893 &  700 &  6.697 $\cdot 10^{-5}$ &  9.698 $\cdot 10^{-6}$ &  7.534 $\cdot 10^{-8}$ & 1.964 $\cdot 10^{-5}$ & 8.337 $\cdot 10^{-4}$ \\
\hline
\end{tabular}
\end{center}
\end{table}

\begin{table}[ht]
\caption*{Table 4 continued}
\begin{center}
\begin{tabular}{| c | c | c | c | c |c |c | c|}
\hline
$M_{A}$ & $M_{H_{S}}$ & $H_{S}\to bb$ & $H_{S}\to \tau\tau$ & $H_{S}\to \gamma\gamma$ &
$H_{S} \to h + h$ &$H_{S}\to tt$\\
\hline
  997 &   60 &  3.686 $\cdot 10^{-3}$ &  3.426 $\cdot 10^{-4}$ &  2.364 $\cdot 10^{-7}$ &  - &  - \\
  997 &   80 &  3.705 $\cdot 10^{-3}$ &  3.646 $\cdot 10^{-4}$ &  1.220 $\cdot 10^{-7}$ &  - &  - \\
  997 &   90 &  3.732 $\cdot 10^{-3}$ &  3.767 $\cdot 10^{-4}$ &  1.560 $\cdot 10^{-7}$ &  - &  - \\
  997 &  100 &  3.695 $\cdot 10^{-3}$ &  3.799 $\cdot 10^{-4}$ &  3.227 $\cdot 10^{-8}$ &  - &  - \\
  996 &  150 &  4.110 $\cdot 10^{-3}$ &  4.583 $\cdot 10^{-4}$ &  9.523 $\cdot 10^{-7}$ &  - &  - \\
  995 &  200 &  3.516 $\cdot 10^{-3}$ &  4.140 $\cdot 10^{-4}$ &  7.459 $\cdot 10^{-7}$ &  - &  - \\ 
  995 &  250 &  3.266 $\cdot 10^{-3}$ &  3.988 $\cdot 10^{-4}$ &  2.122 $\cdot 10^{-6}$ &  2.152 $\cdot 10^{-4}$ &  - \\
  995 &  300 &  3.014 $\cdot 10^{-3}$ &  3.796 $\cdot 10^{-4}$ &  2.608 $\cdot 10^{-6}$ &  4.453 $\cdot 10^{-4}$ &  - \\
  994 &  400 &  7.315 $\cdot 10^{-4}$ &  9.680 $\cdot 10^{-5}$ &  8.071 $\cdot 10^{-7}$ &  5.161 $\cdot 10^{-6}$ &  2.804 $\cdot 10^{-3}$ \\
  994 &  500 &  3.426 $\cdot 10^{-4}$ &  4.700 $\cdot 10^{-5}$ &  5.456 $\cdot 10^{-7}$ &  1.176 $\cdot 10^{-5}$ &  2.750 $\cdot 10^{-3}$ \\
  994 &  600 &  2.852 $\cdot 10^{-4}$ &  4.028 $\cdot 10^{-5}$ &  6.339 $\cdot 10^{-7}$ &  1.067 $\cdot 10^{-5}$ &  2.312 $\cdot 10^{-3}$ \\
  994 &  700 &  1.484 $\cdot 10^{-4}$ &  2.148 $\cdot 10^{-5}$ &  2.816 $\cdot 10^{-7}$ &  2.650 $\cdot 10^{-4}$ &  1.411 $\cdot 10^{-3}$ \\
  990 &  800 &  1.916 $\cdot 10^{-5}$ &  2.832 $\cdot 10^{-6}$ &  2.770 $\cdot 10^{-8}$ &  2.969 $\cdot 10^{-7}$ &  3.705 $\cdot 10^{-4}$ \\
\hline
 1198 &   60 &  1.042 $\cdot 10^{-3}$ &  9.694 $\cdot 10^{-5}$ &  9.963 $\cdot 10^{-8}$ &  - &  - \\
 1197 &   80 &  1.048 $\cdot 10^{-3}$ &  1.033 $\cdot 10^{-4}$ &  4.304 $\cdot 10^{-8}$ &  - &  - \\
 1197 &  100 &  1.034 $\cdot 10^{-3}$ &  1.067 $\cdot 10^{-4}$ &  7.044 $\cdot 10^{-8}$ &  - &  - \\
 1197 &  150 &  1.006 $\cdot 10^{-3}$ &  1.122 $\cdot 10^{-4}$ &  2.134 $\cdot 10^{-7}$ &  - &  - \\
 1196 &  200 &  9.342 $\cdot 10^{-4}$ &  1.099 $\cdot 10^{-4}$ &  2.441 $\cdot 10^{-7}$ &  - &  - \\ 
 1196 &  300 &  9.621 $\cdot 10^{-4}$ &  1.213 $\cdot 10^{-4}$ &  8.096 $\cdot 10^{-7}$ &  1.083 $\cdot 10^{-4}$ &  - \\
 1196 &  400 &  2.339 $\cdot 10^{-4}$ &  3.092 $\cdot 10^{-5}$ &  2.908 $\cdot 10^{-7}$ &  2.944 $\cdot 10^{-5}$ &  7.542 $\cdot 10^{-4}$ \\
 1195 &  500 &  1.178 $\cdot 10^{-4}$ &  1.615 $\cdot 10^{-5}$ &  2.274 $\cdot 10^{-7}$ &  1.641 $\cdot 10^{-5}$ &  8.261 $\cdot 10^{-4}$ \\
 1195 &  600 &  8.381 $\cdot 10^{-5}$ &  1.183 $\cdot 10^{-5}$ &  2.198 $\cdot 10^{-7}$ &  2.218 $\cdot 10^{-5}$ &  8.535 $\cdot 10^{-4}$ \\
 1195 &  700 &  6.596 $\cdot 10^{-5}$ &  9.539 $\cdot 10^{-6}$ &  2.034 $\cdot 10^{-7}$ &  1.101 $\cdot 10^{-6}$ &  8.477 $\cdot 10^{-4}$ \\
 1194 &  800 &  4.805 $\cdot 10^{-5}$ &  7.096 $\cdot 10^{-6}$ &  1.575 $\cdot 10^{-7}$ &  5.938 $\cdot 10^{-6}$ &  8.570 $\cdot 10^{-4}$ \\
\hline
 1398 &   60 &  3.346 $\cdot 10^{-4}$ &  3.116 $\cdot 10^{-5}$ &  4.626 $\cdot 10^{-8}$ &  - &  - \\
 1398 &   80 &  3.398 $\cdot 10^{-4}$ &  3.348 $\cdot 10^{-5}$ &  1.272 $\cdot 10^{-8}$ &  - &  - \\
 1398 &  100 &  3.244 $\cdot 10^{-4}$ &  3.347 $\cdot 10^{-5}$ &  2.164 $\cdot 10^{-8}$ &  - &  - \\
 1398 &  150 &  3.163 $\cdot 10^{-4}$ &  3.529 $\cdot 10^{-5}$ &  6.580 $\cdot 10^{-8}$ &  - &  - \\
 1396 &  200 &  2.557 $\cdot 10^{-4}$ &  3.008 $\cdot 10^{-5}$ &  6.910 $\cdot 10^{-8}$ &  - &  - \\ 
 1397 &  300 &  2.768 $\cdot 10^{-4}$ &  3.490 $\cdot 10^{-5}$ &  2.365 $\cdot 10^{-7}$ &  5.294 $\cdot 10^{-5}$ &  - \\
 1396 &  400 &  8.693 $\cdot 10^{-5}$ &  1.149 $\cdot 10^{-5}$ &  1.182 $\cdot 10^{-7}$ &  1.380 $\cdot 10^{-5}$ &  2.130 $\cdot 10^{-4}$ \\
 1396 &  500 &  4.476 $\cdot 10^{-5}$ &  6.133 $\cdot 10^{-6}$ &  9.791 $\cdot 10^{-8}$ &  7.447 $\cdot 10^{-6}$ &  2.522 $\cdot 10^{-4}$ \\
 1396 &  600 &  3.327 $\cdot 10^{-5}$ &  4.693 $\cdot 10^{-6}$ &  1.051 $\cdot 10^{-7}$ &  1.172 $\cdot 10^{-5}$ &  2.869 $\cdot 10^{-4}$ \\
 1397 &  700 &  1.939 $\cdot 10^{-5}$ &  2.801 $\cdot 10^{-6}$ &  7.272 $\cdot 10^{-8}$ &  2.656 $\cdot 10^{-5}$ &  1.125 $\cdot 10^{-4}$ \\
 1397 &  800 &  1.066 $\cdot 10^{-5}$ &  1.570 $\cdot 10^{-6}$ &  4.274 $\cdot 10^{-8}$ &  4.312 $\cdot 10^{-5}$ &  5.132 $\cdot 10^{-5}$ \\
\hline
\end{tabular}
\end{center}
\end{table}

\begin{table}[ht]
\caption*{Table 4 continued}
\begin{center}
\begin{tabular}{| c | c | c | c | c |c |c | c|}
\hline
$M_{A}$ & $M_{H_{S}}$ & $H_{S}\to bb$ & $H_{S}\to \tau\tau$ & $H_{S}\to \gamma\gamma$ &
$H_{S} \to h + h$ &$H_{S}\to tt$\\
\hline
 1598 &   60 &  9.977 $\cdot 10^{-5}$ &  9.270 $\cdot 10^{-6}$ &  1.654 $\cdot 10^{-8}$ &  - &  - \\
 1598 &   80 &  1.005 $\cdot 10^{-4}$ &  9.886 $\cdot 10^{-6}$ &  7.958$\cdot 10^{-10}$ &  - &  - \\
 1599 &  100 &  9.688 $\cdot 10^{-5}$ &  9.949 $\cdot 10^{-6}$ &  7.944 $\cdot 10^{-9}$ &  - &  - \\
 1598 &  150 &  8.955 $\cdot 10^{-5}$ &  9.981 $\cdot 10^{-6}$ &  2.019 $\cdot 10^{-8}$ &  - &  - \\
 1597 &  200 &  8.357 $\cdot 10^{-5}$ &  9.829 $\cdot 10^{-6}$ &  2.362 $\cdot 10^{-8}$ &  - &  - \\
 1597 &  300 &  6.934 $\cdot 10^{-5}$ &  8.742 $\cdot 10^{-6}$ &  5.935 $\cdot 10^{-8}$ &  2.144 $\cdot 10^{-5}$ &  - \\
 1597 &  400 &  2.690 $\cdot 10^{-5}$ &  3.550 $\cdot 10^{-6}$ &  3.879 $\cdot 10^{-8}$ &  4.499 $\cdot 10^{-6}$ &  5.207 $\cdot 10^{-5}$ \\
 1597 &  500 &  1.381 $\cdot 10^{-5}$ &  1.892 $\cdot 10^{-6}$ &  3.258 $\cdot 10^{-8}$ &  3.516 $\cdot 10^{-6}$ &  6.507 $\cdot 10^{-5}$ \\
 1597 &  600 &  9.845 $\cdot 10^{-6}$ &  1.388 $\cdot 10^{-6}$ &  3.436 $\cdot 10^{-8}$ &  7.181 $\cdot 10^{-6}$ &  7.156 $\cdot 10^{-5}$ \\
 1598 &  700 &  3.917 $\cdot 10^{-6}$ &  5.651 $\cdot 10^{-7}$ &  1.631 $\cdot 10^{-8}$ &  5.658 $\cdot 10^{-6}$ &  9.579 $\cdot 10^{-6}$ \\
 1598 &  800 &  2.307 $\cdot 10^{-6}$ &  3.394 $\cdot 10^{-7}$ &  1.107 $\cdot 10^{-8}$ &  1.439 $\cdot 10^{-5}$ &  4.832 $\cdot 10^{-6}$ \\
\hline
 1798 &   60 &  3.466 $\cdot 10^{-5}$ &  3.223 $\cdot 10^{-6}$ &  7.182 $\cdot 10^{-9}$ &  - &  - \\
 1798 &   80 &  3.497 $\cdot 10^{-5}$ &  3.448 $\cdot 10^{-6}$ &  1.673 $\cdot 10^{-9}$ &  - &  - \\
 1798 &  100 &  3.470 $\cdot 10^{-5}$ &  3.576 $\cdot 10^{-6}$ &  2.998 $\cdot 10^{-9}$ &  - &  - \\
 1798 &  150 &  3.137 $\cdot 10^{-5}$ &  3.497 $\cdot 10^{-6}$ &  5.759 $\cdot 10^{-9}$ &  - &  - \\
 1797 &  200 &  2.989 $\cdot 10^{-5}$ &  3.515 $\cdot 10^{-6}$ &  8.838 $\cdot 10^{-9}$ &  - &  - \\
 1798 &  300 &  1.832 $\cdot 10^{-5}$ &  2.310 $\cdot 10^{-6}$ &  1.554 $\cdot 10^{-8}$ &  1.116 $\cdot 10^{-5}$ &  - \\
 1797 &  400 &  1.062 $\cdot 10^{-5}$ &  1.403 $\cdot 10^{-6}$ &  1.567 $\cdot 10^{-8}$ &  1.758 $\cdot 10^{-6}$ &  1.693 $\cdot 10^{-5}$ \\
 1797 &  500 &  5.741 $\cdot 10^{-6}$ &  7.865 $\cdot 10^{-7}$ &  1.414 $\cdot 10^{-8}$ &  1.626 $\cdot 10^{-7}$ &  2.344 $\cdot 10^{-5}$ \\
 1797 &  600 &  4.217 $\cdot 10^{-6}$ &  5.944 $\cdot 10^{-7}$ &  1.569 $\cdot 10^{-8}$ &  3.764 $\cdot 10^{-8}$ &  2.627 $\cdot 10^{-5}$ \\
 1798 &  700 &  9.225 $\cdot 10^{-7}$ &  1.331 $\cdot 10^{-7}$ &  4.140 $\cdot 10^{-9}$ &  8.232 $\cdot 10^{-6}$ &  7.913 $\cdot 10^{-7}$ \\
 1798 &  800 &  5.024 $\cdot 10^{-7}$ &  7.397 $\cdot 10^{-8}$ &  2.638 $\cdot 10^{-9}$ &  1.011 $\cdot 10^{-5}$ &  1.752 $\cdot 10^{-7}$ \\
\hline
 1999 &   60 &  1.360 $\cdot 10^{-5}$ &  1.265 $\cdot 10^{-6}$ &  2.502 $\cdot 10^{-9}$ &  - &  - \\
 1998 &   80 &  1.366 $\cdot 10^{-5}$ &  1.342 $\cdot 10^{-6}$ &  7.307$\cdot 10^{-11}$ &  - &  - \\
 1998 &  100 &  1.353 $\cdot 10^{-5}$ &  1.393 $\cdot 10^{-6}$ &  3.954$\cdot 10^{-10}$ &  - &  - \\
 1998 &  150 &  1.388 $\cdot 10^{-5}$ &  1.549 $\cdot 10^{-6}$ &  2.657 $\cdot 10^{-9}$ &  - &  - \\
 1998 &  200 &  1.351 $\cdot 10^{-5}$ &  1.588 $\cdot 10^{-6}$ &  5.393 $\cdot 10^{-9}$ &  - &  - \\
 1998 &  300 &  3.173 $\cdot 10^{-6}$ &  3.999 $\cdot 10^{-7}$ &  2.558 $\cdot 10^{-9}$ &  7.529 $\cdot 10^{-6}$ &  - \\
 1998 &  400 &  4.428 $\cdot 10^{-6}$ &  5.854 $\cdot 10^{-7}$ &  6.683 $\cdot 10^{-9}$ &  1.123 $\cdot 10^{-6}$ &  5.759 $\cdot 10^{-6}$ \\
 1997 &  500 &  2.618 $\cdot 10^{-6}$ &  3.588 $\cdot 10^{-7}$ &  6.633 $\cdot 10^{-9}$ &  5.120$\cdot 10^{-10}$ &  9.458 $\cdot 10^{-6}$ \\
 1997 &  600 &  1.941 $\cdot 10^{-6}$ &  2.738 $\cdot 10^{-7}$ &  7.489 $\cdot 10^{-9}$ &  2.722 $\cdot 10^{-7}$ &  1.050 $\cdot 10^{-5}$ \\
 1998 &  700 &  1.153 $\cdot 10^{-6}$ &  1.665 $\cdot 10^{-7}$ &  5.945 $\cdot 10^{-9}$ &  2.699 $\cdot 10^{-6}$ &  4.700 $\cdot 10^{-6}$ \\
 1998 &  800 &  3.013 $\cdot 10^{-7}$ &  4.442 $\cdot 10^{-8}$ &  1.803 $\cdot 10^{-9}$ &  4.388 $\cdot 10^{-6}$ &  2.682 $\cdot 10^{-7}$ \\
\hline
\end{tabular}
\end{center}
\end{table}

\begin{table}[ht]
\caption{Possible cross sections $\sigma$ at 13~TeV for $ggF\to A \to Z + h$}
\begin{center}
\begin{tabular}{| c | c | c | c | c |c |c | c|}
\hline
$M_{A}$ [GeV] & 400 & 500 & 600 & 700 & 800 & 900 \\
\hline
$\sigma$ [fb] & 25.9 & 30.8 & 8.08 & 4.72 & 2.12 & 1.29 \\
\hline
$M_{A}$ [GeV] & 1000 & 1200 & 1400 & 1600 & 1800 & 2000 \\
\hline
$\sigma$ [fb] & 5.76$\cdot 10^{-1}$ & 1.93$\cdot 10^{-1}$ & 4.83$\cdot 10^{-2}$ & 1.37$\cdot 10^{-2}$ & 4.69$\cdot 10^{-3}$ & 2.10$\cdot 10^{-3}$ \\
\hline

\end{tabular}
\end{center}
\end{table}

\end{document}